\documentclass[12pt]{article}
\pdfoutput=1
\usepackage{draft}
\newtheorem{theorem}{Theorem}
\newtheorem{corollary}{Corollary}

\usepackage{colortbl}
\usepackage{relsize}

\usepackage{tikz-cd}
\usetikzlibrary{arrows,snakes,shapes.arrows,decorations.markings}
     \tikzset{>=triangle 90}
     \tikzstyle{bbc}=[draw,circle,fill=black,scale=.75]
     \tikzstyle{rc}=[circle,fill=red,scale=.6]
     \tikzstyle{wc}=[draw,circle,scale=.75]

\renewcommand{\D}{\text{D}}
\renewcommand{\O}{\text{O}}
\newcommand{\SO}{\text{SO}}
\newcommand{\Spin}{\text{Spin}}
\renewcommand{\D}{\text{D}}
\renewcommand{\S}{\text{S}}

\newcommand{\cen}{\text{C}}

\newcommand{\CC}{\mathrm{C}}
\newcommand{\Co}{\mathrm{Co}}
\newcommand{\U}{\mathrm{U}}
\renewcommand{\H}{\mathrm{H}}
\renewcommand{\A}{\text{A}}
\renewcommand{\B}{\text{B}}

\newcommand{\bN}{\mathbb{N}}
\newcommand{\mygcd}[1]{\gcd(#1)}
\newcommand{\dv}{\,|\,}
\newcommand{\bdv}{\,\bigg|\,}

\newcommand{\sgn}{\text{sgn}}

\newcommand{\T}{\mathcal{T}}

\newcommand{\Duntw}{{V^{f\natural}_\text{tw}}}
\newcommand{\DunArf}{{\widetilde{V}^{f\natural}\,}}
\newcommand{\Dun}{{V^{f\natural}}}
\newcommand{\DunFull}{{V^{f\natural}_\text{NS} \oplus V^{f\natural}_\text{R}}}
\newcommand{\DunS}{{V^{s\natural}}}

\newcommand{\ch}{\chi}
\newcommand{\wi}{\cI}
\newcommand{\eg}{Z}
\newcommand{\EG}{\cZ}
\newcommand{\Hilb}{\mathcal H}

\newcommand{\myfrac}[2]{\frac{#1}{#2}}
\newcommand{\R}{\mathrm{R}}
\newcommand{\NS}{\mathrm{NS}}
\newcommand{\ER}{E_\R}

\def\half{\frac12}
\def\ZZ{{\mathbb Z}}

\def\NN{{\mathbb N}}

\def\cT{{\mathcal T}}
\def\sfT{{\mathsf T}}

\def\no{\nonumber}
\def\text{\mathrm}
\usepackage{tikzsymbols}
\usepackage{halloweenmath}
\usepackage{lipsum} 

\newcommand{\Z}{\bZ}

\begin{document}

\begin{titlepage}

\begin{center}

\hfill \\
\hfill \\
\vskip 1cm

\title{Topological modularity of Supermoonshine}

\author{Jan Albert,$^{1,2}$ Justin Kaidi,$^{3,2,4}$ Ying-Hsuan Lin$^{5}$}
  
\address{${}^{1}$C. N. Yang Institute for Theoretical Physics,\\  Stony Brook University, Stony Brook, NY 11794-3840, USA\\

${}^{2}$Simons Center for Geometry and Physics, \\ Stony Brook University, Stony Brook, NY 11794-3636, USA\\

${}^{3}$Department of Physics, University of Washington, Seattle, WA, 98195, USA\\

${}^{4}$Kavli Institute for the Physics and Mathematics of the Universe, \\
  University of Tokyo, Kashiwa, Chiba 277-8583, Japan\\

${}^{5}$Jefferson Physical Laboratory, Harvard University, Cambridge, MA 02138, USA}

\email{jan.albertiglesias@stonybrook.edu, jkaidi@uw.edu, yhlin@fas.harvard.edu}

\end{center}
\vspace{0.3 in}

\begin{abstract}
   The theory of topological modular forms (TMF) predicts that elliptic genera of physical theories satisfy a certain divisibility property, determined by the theory's gravitational anomaly. In this note we verify this prediction in Duncan's Supermoonshine module, as well as in tensor products and orbifolds thereof. Along the way we develop machinery for computing the elliptic genera of general alternating orbifolds and discuss the relation of this construction to the elusive ``periodicity class" of TMF.
\end{abstract}

\vfill

\end{titlepage}

\tableofcontents

\section{Introduction and summary}

Moonshine \cite{Conway:1979qga} is one of the most beautiful subjects at the interface of mathematics, physics, and folklore.  What originated from a curious observation about modular forms and the Monster group has transcended into the fields of conformal field theory \cite{frenkel1984natural}, string theory \cite{Paquette:2016xoo}, and quantum gravity \cite{W}. Subsequent developments have lead to the discovery of Supermoonshine \cite{duncan2007super,duncan2015moonshine}, Mathieu Moonshine \cite{Eguchi:2010ej,Gannon:2012ck,Gaberdiel:2012um,Gaberdiel:2011fg,Taormina:2011rr,Banlaki:2018pcc}, and Umbral Moonshine \cite{Cheng:2012tq,Cheng:2013wca,Duncan:2015rga,Duncan:2015xoa,Cheng:2014zpa,Kachru:2016ttg}.  

In a similar spirit, topological modular forms \cite{douglas2014topological} have begun to make a surprise appearance in physics thanks to a conjecture by Stolz and Teichner \cite{ST1,ST2} based on earlier work by Segal \cite{segal1987elliptic,Segal}.  The conjecture roughly states the following:
\begin{enumerate}
    \item Every 2d supersymmetric quantum field theory (SQFT) with $\cN = (0,1)$ supersymmetry can be associated with a ``topological modular form,'' or more precisely a class in TMF,
    \item Every class in TMF can be realized by at least one $\cN = (0,1)$ SQFT,
    \item TMF is a \emph{complete} supersymmetric deformation invariant, i.e.\ any two SQFTs can be continuously connected \emph{if and only} if they are associated with the same topological modular forms.
\end{enumerate}
Although the map between SQFTs and TMF is not yet fully understood (though see \cite{Gaiotto:2019asa,Yonekura:2022reu} for key progress), in some cases the image of the map is a familiar object: the elliptic genus, i.e.\ the torus partition function with Ramond boundary conditions along both space and time.  The non-surjectivity of the map from SQFTs to the space of modular forms implies a remarkable divisibility property of certain coefficients in the elliptic genus, as will be reviewed below for the cases of interest to us.\footnote{The physicist readers are referred to \cite{Tachikawa:2021mvw,Lin:2021bcp} for a friendly introduction to this divisibility.
}

By now, intricate connections between topological modular forms and Moonshine have been uncovered \cite{Gaiotto:2018ypj,Johnson-Freyd:2020itv,Theo,Lin:2022wpx}, and this note continues this pursuit.
The protagonist of our story is Duncan's Supermoonshine module $\Dun$ \cite{duncan2007super}, a holomorphic $\cN=1$ supersymmetric conformal field theory (SCFT) with central charge $c=12$ that enjoys Conway symmetry.\footnote{This theory actually has $\cN = (1,1)$ supersymmetry because the anti-holomorphic sector can be equipped with trivial supersymmetry.
}\footnote{In the math literature, the notation $\Dun$ only refers to the supersymmetric vertex operator algebra (SVOA) of the Neveu-Schwarz sector, whereas the Ramond sector is denoted by $\Duntw$.  In this note, we slightly abuse $\Dun$ to mean the entire fermionic theory $\Dun \oplus \Duntw$.
}
Its (twisted and twined) elliptic genera are all constants due to supersymmetry, while its partition functions with other boundary conditions exhibit many of the same extraordinary properties as their Monster cousins, including the celebrated genus-zero property \cite{duncan2007super,duncan2015moonshine,Harrison:2018joy,Harrison:2021gnp}. A brief review of the construction and properties of $\Dun$ is given in Section~\ref{Sec:Duncan}.

For a general $c = 12n$ holomorphic $\cN=1$ SCFT whose elliptic genus is a constant and equal to the Witten index $\wi$, the aforementioned divisibility property states that\footnote{
  Let the prime factorization of a triple of natural numbers $D, n, a$ be
    \ie
        D = \prod_i p_i^{\delta_i}, \quad n = \prod_i p_i^{\nu_i}, \quad a = \prod_i p_i^{\alpha_i}, \quad \delta_i, \nu_i, \alpha_i \in \mathbb{Z}_{\ge 0}~.
    \fe
    Then 
    \ie
        \frac{D}{\gcd(D,n)}\, \bdv \, a~ \quad\Leftrightarrow\quad \delta_i \le \min(\delta_i, \nu_i) + \alpha_i \quad \forall i~,
    \fe
    and
    \ie
        D \dv n \, a \quad\Leftrightarrow\quad \delta_i \le \nu_i + \alpha_i \quad \forall i~.
    \fe
    If $\delta_i \le \nu_i$, then both conditions are obviously true; if $\delta_i \ge \nu_i$, then the two conditions become identical.  In \eqref{Divisibility}, $D = 24$ and $a = \cI$.
    \label{footnote:1}
}
\ie\label{Divisibility}
    \frac{24}{\gcd(24,n)}\, \bdv \,\wi \,\,\quad \text{or\,\,equivalently}\,\,\quad 24 \dv n \, \wi~.
\fe
The Supermoonshine module has precisely $\cI = -24$, saturating divisibility for $n=1$. Since this divisibility is at present still conjectural, it is a valuable exercise to check its validity in a variety of theories of physical interest. Such checks were performed in \cite{Lin:2022wpx} in the context of the Monster module $V^\natural$ as well as its tensor products and various orbifolds. Conversely, assuming the validity of the conjecture, one can rule out the existence of a number of tentative VOAs proposed in the literature, including many of the extremal CFTs proposed in \cite{W}.

In the current work, we perform a similar exercise for the Supermoonshine module $\Dun$. In particular, we check the validity of the divisibility criterion for tensor products of $\Dun$, together with orbifolds by $\S_n$ and $\A_n$ permutation symmetries. We also allow for orbifolds by non-anomalous cyclic subgroups of the diagonal $\mathrm{Co}_0$ symmetry.  While primarily serving as a check of the Stolz-Teichner conjecture, this exercise has an important secondary motivation: namely, to develop the tools necessary for realizing a special class in TMF containing the ``periodicity elements,'' whose definition we now review.

\subsection{Periodicity elements}

TMF is a generalized cohomology ring graded by an integer $\nu$, where the multiplication and addition operations correspond physically to taking the tensor product and direct sum of SQFTs and $\nu$ characterizes the gravitational anomaly.\footnote{The gravitational anomaly $\nu$ is conventionally normalized such that a chiral fermion has $\nu = 1$.
}
For SCFTs, the quantity $\nu$ is related to the chiral central charge by $\nu = 2(c_\text{R} - c_\text{L})$, so that in particular a holomorphic  SCFT with $(c_\text{L},c_\text{R} )= (c,0)$ has $\nu = -2c$. Intuitively, the group TMF$_\nu$ captures how much data beyond the gravitational anomaly $\nu$ is necessary to specify the deformation class of an SQFT. The notion of deformation class here includes, but is not necessarily limited to, the identification of all theories connected by RG flows (induced by either relevant deformations or vacuum expectation values) as well as theories connected by marginal deformations \cite{Gaiotto:2019asa}.

It is known that the cohomology ring has periodicity $\nu \sim \nu + 576$. This is a rather remarkable property: it means that the set of deformation classes of SQFTs with gravitational anomaly $\nu$ is identical to that of SQFTs with gravitational anomaly $\nu + 576$. Indeed, there exists a special class in TMF$_{-576}$ called the ``periodicity class'' such that every class of TMF$_{\nu - 576}$ is obtained from a unique class in TMF$_{\nu}$ by taking the product with the periodicity class. An SQFT with $\nu = -576$ realizes an element in the periodicity class if and only if its elliptic genus is a constant with value $\pm 1$.\footnote{One can also realize the periodicity class in TMF$_{576}$, but this is less interesting since it can be realized by an $\cN=(0,1)$ sigma model, as described briefly later.  Furthermore,
 given a holomorphic $\cN=1$ theory realizing $\nu = -576$ (which will be our focus below), we can get another $\n=+576$ theory by exchanging left- and right-movers.}

Because a constant elliptic genus is a highly non-generic feature in systems without supersymmetry, a natural starting point for realizing an element in the periodicity class is to consider holomorphic $\cN=1$ SCFTs with central charge $c = 288$. 
This leads us to the study of theories constructed from $\Dun$.
Indeed, according to \cite[Example 2.4.1]{Johnson-Freyd:2017ble}, the 't Hooft anomaly of the $\Co_1$ symmetry of $\Dun$ realizes the generator of 
$\S\H^3(\Co_1) = \bZ_{24}$.\footnote{ We remind the reader that the supercohomology group $\text{SH}^d(G)$ comprises the first three layers of the Atiyah–Hirzebruch spectral sequence for the spin bordism group $\Omega^{\mathrm{Spin}}_d(BG)$ \cite{wang2018towards,Gaiotto:2017zba}. As a set, this is equivalent to $\text{H}^d(G,\U(1))\oplus \text{H}^{d-1}(G, \ZZ_2) \oplus \text{H}^{d-2}(G,\ZZ_2)$--- referred to as the bosonic, Gu-Wen, and Majorana layers, respectively---on the $E_2$ page, and is generally reduced by non-trivial differentials on the higher pages. For $d=3$ the group $\text{SH}^3(G)$ is identical to $\Omega^{\mathrm{Spin}}_3(G)$ \cite{brumfiel2016pontrjagin}, but for larger $d$ it captures less information.}

Hence, the diagonal $\Co_1$ symmetry of $(\Dun)^{\otimes n}$ is non-anomalous when $24 \dv n$.
For $n=24$, the chiral central charge $c=12 \times 24 = 288$ gives precisely the amount of gravitational anomaly needed for the periodicity class of TMF, and one may then hope (on the basis of aesthetics alone) that one periodicity element is realized by the theory $\Dun^{\otimes 24}/\Co_1$. 

Unfortunately, with present technology, it is not possible to conclusively refute or confirm this guess. Indeed, ignorance of the generalized McKay-Thompson data for $\Dun$ prevents one from computing the full Witten index of the $\Co_1$ orbifold. Nevertheless, there is evidence to suggest that this first guess is incorrect. In particular, the Witten index of $\Dun^{\otimes 24}$ is $24^{24}$, whereas 
$\Co_1$ ``only'' has 4,157,776,806,543,360,000 
elements, 15 orders of magnitude smaller. It thus seems extremely unlikely---though not strictly speaking impossible---for the $\Co_1$ orbifold to have Witten index $\pm 1$. One is led to consider alternative constructions. 

One closely related construction is to allow for permutation orbifolds, e.g.\ 
$\Dun^{\otimes 24}/\S_{24}$. Indeed, permutation orbifolds are well-known to give rise to massive reductions in the index or the degeneracies in the light spectra \cite{Haehl:2014yla,Keller:2017rtk}, and thus seem well-suited for the current task. Lo and behold, allowing for permutation orbifolds \textit{does} enable one to identify a periodicity element. To see this, we note that the Witten index of $\Dun^{\otimes 24}$ is $24^{24}$, whereas that of $\Dun^{\otimes 24}/\S_{24}$ is $-25,499,225$ (the machinery for performing the latter computation will be introduced in the main text). A nice fact is that these two numbers are coprime. Given any two coprime integers $m$ and $n$, Bezout's identity ensures that there exist integers $x$ and $y$ such that $m x + n y = 1$. Solving for the appropriate Bezout pair, we find that 
\bea
\label{PeriodicityTheory}
    24697376 \times \Dun^{\otimes 24} \oplus 1291795102224619090515486568295959 \times \Dun^{\otimes 24}/\S_{24}~
\eea
has Witten index 1, and hence is an element of the periodicity class.\footnote{According to \cite{Theo}, this fact was first found by Gaiotto in unpublished work.} While this may be viewed as a success, it certainly leaves something to be desired. In particular, the theory constructed above has massively degenerate vacua, even at finite volume. It would be more satisfying to identify a periodicity element with a unique vacuum, assuming that such a theory exists. 

The original statement by Stolz and Teichner about the SQFT realizability of TMF did not require the SQFT to be ``indecomposable'', namely to have a unique vacuum on a spatial circle of finite size with Neveu-Schwarz boundary conditions.\footnote{A decomposable QFT is one that can be written as a direct sum where each summand is a superselection sector, or more precisely a ``universe'' \cite{Hellerman:2006zs,Tanizaki:2019rbk,Aminov:2019hwg,Sharpe:2019ddn,Komargodski:2020mxz}.  In terms of local operators, each direct summand gives rise to a topological point operator (a projector onto the summand) generating a top-form symmetry.  If the SQFT is quantized on a spatial circle of finite size, then each direct summand gives rise to an \emph{exact} vacuum.  This notion of vacuum degeneracy is different from the usual one in Minkowski space, or equivalently, on a spatial circle of \emph{infinite} size.  A study of the Minkowski vacuum degeneracy of the supersymmetric three-sphere sigma model using TMF can be found in \cite{Gaiotto:2019asa,Gaiotto:2019gef}.
}
However, there is reason to believe that every class in TMF is realizable by an indecomposable SQFT.\footnote{It is also interesting to ask whether every TMF class can be realized by a \textit{conformal} field theory. For theories with a non-vanishing gravitational anomaly the infrared of an SQFT is expected to be an SCFT, so this seems reasonable, though even some of the simplest classes have yet to be realized \cite{GPPV}.  One may further ask about realizability via indecomposable CFTs; if there were a TMF class that could not be realized by an indecomposable CFT, then this would necessitate degenerate vacua (in Minkowski space).
}
For example, every TMF class with $-24 \le \nu \le 24$ has been realized by such an SQFT \cite{GPPV}. Moreover, it is known \cite{devalapurkar2019ando} that every tmf class can be realized by a connected string manifold, which serves as the target space of an indecomposable $\cN = (0,1)$ sigma model SQFT \cite{Moore:1984dc,Moore:1984ws,Manohar:1984zj,Witten:1985mj}.
Since TMF is obtained by adjoining tmf with the periodicity class, the realizability via indecomposable
theories 
would be true in general if it is true for the periodicity class.

To obtain an indecomposable periodicity element, 
we may now try to combine the symmetric orbifold with an orbifold by $\Co_0$ or $\Co_1$. Beginning with $\Dun^{\otimes n}/\S_n \times \Co_0$, we quickly see that this cannot do the job. Indeed, though the theory (including all spin structures)
has an action by  $\Co_0 = 2.\Co_1$, the $\bZ_2$ acts trivially on the Neveu-Schwarz sector of $\Dun^{\otimes n}/\S_n$. Thus the gauged theory $\Dun^{\otimes n}/\S_n \times \Co_0$ will \emph{always} have degenerate vacua in the Neveu-Schwarz sector, even at finite volume. If we are going to get an indecomposable theory, it would seem that we only want to gauge $\Co_1$.

However, it does not actually make sense to gauge $\Dun^{\otimes n}/\S_n$ by $\Co_1$ due to a certain mixed anomaly. To phrase this, it is useful to split the anomaly of $\Dun$ as $ \S\H^3(\Co_1) = \H^3(\Co_1, U(1)). \H^2(\Co_1, \ZZ_2) = \ZZ_{12}. \ZZ_2$, where the second piece corresponds to a projective phase for $\Co_1$ in the R sector. The fact that the $\Co_1$ symmetry is realized projectively in the R sector can alternatively be phrased as saying that the R sector is in a linear representation of the (non-split) central extension of $\Co_1$ by $\ZZ_2$, with extension class specified by the anomaly. This is none other than $\Co_0 = 2 . \Co_1$. So the fact that the R sector transforms faithfully under $\Co_0$ is a signature of the anomaly, and this signature turns out to be present in $\Dun^{\otimes n}/\S_n$ as well. 
On the other hand,  as we will see the $\Co_0$ action can be unfaithful for $\Dun^{\otimes n}/\A_n$ and $n$ even, where $\A_n$ is the alternating group (which is a $\ZZ_2$ quotient of $\S_n$).

We are thus finally led to consider $\Dun^{\otimes 24}/\A_{24} \times \Co_1$ as a candidate for an indecomposable periodicity element. As a first check, the Witten index of $\Dun^{\otimes 24}/\A_{24} $ can be easily computed using formulae given in the main text and is smaller than the order of $\Co_1$ by 2-3 orders of magnitude (c.f. Table \ref{Tab:24}), making it conceivable that the Witten index of $\Dun^{\otimes 24}/\A_{24} \times \Co_1$ be $\pm1$. Of course, there remain many variants on the theme---in particular, we could allow for gaugings with discrete torsion in\footnote{Here $\mho_{\mathrm{Spin}}$ denotes the Pontryagin dual of spin bordism, $\mho_{\text{Spin}}^d(BG) = \mathrm{Hom}(\Omega^{\text{Spin}}_d(BG),\U(1))$. We could also allow for discrete torsion involving $\Co_1$, but will not do so here.}
\bea
\mho^2_{\mathrm{Spin}}(\text{B}\A_{24}) \cong  \mho^2_{\mathrm{Spin}}(pt) \oplus \H^2(\A_{24},\U(1))~. 
\eea 
The generator of the first group on the right-hand side is the invertible field theory $(-1)^{\mathrm{Arf}}$.\footnote{The invertible field theory $(-1)^{\mathrm{Arf}}$ arises in the IR limit of the Kitaev chain \cite{Kitaev:2000nmw}.  On the torus, it is $-1$ for the Ramond-Ramond (non-bounding) spin structure and $+1$ otherwise.} The generator of the second is a certain 2-cycle discussed more below. In general, we will use the notation $\widetilde{V}^{f \natural} : = \Dun \otimes (-1)^{\mathrm{Arf}}$, and denote the gauging of $\A_{24}$ with the group cohomology twist by $\cT/ \A^{\mathrm{tor}}_n$, where $\T$ is either $\Dun$ or $\DunArf$. There are then seemingly three alternative gaugings $({\widetilde{V}}^{f\natural})^{\otimes 24}/\A_{24} \times \Co_1$, $\Dun^{\otimes 24}/\A^\text{tor}_{24} \times \Co_1$, and $({\widetilde{V}}^{f\natural})^{\otimes 24}/\A^\text{tor}_{24} \times \Co_1$. 
In fact, we will see that the turning on discrete torsion in $\H^2(\A_{24},\U(1))$ is almost equivalent to taking the tensor product of the seed theory with $(-1)^{\mathrm{Arf}}$, and hence there are really only two distinct cases to consider.\footnote{This is not the case for $\S_n$, where discrete torsion has a more drastic effect.}
Until the relevant data about the generalized McKay-Thompson data for $\Dun$ is obtained, the question of which, if either, gives an indecomposable periodicity element will remain out of reach.\footnote{
    Since $\Dun$ is the $\bZ_2$ orbifold of 24 free Majorana-Weyl fermions, where $\Co_0$ is the centralizer of $\bZ_2$ in $\O(24)$, the twisted and twined elliptic genera can be computed quite straightforwardly in the free fermion description.  However, a technical first step is to figure out the conjugacy classes of commuting pairs of elements of $\Co_0$, i.e.\ the set of pairs $(g, h) \in \Co_0 \times \Co_0$ modulo $(g, h) \sim (fgf^{-1}, fhf^{-1})$ for all $f \in \Co_0$.
    We thank Theo Johnson-Freyd and an anonymous PTEP referee for comments on this point.
}

\begin{table}
    \centering
    \begin{tabular}{|c|c|}
        \hline
        Permutation orbifold & Witten index
        \\\hline
        $\Dun^{\otimes 24}/\S_{24}$ & $-$25,499,225
        \\
        $\DunArf^{\otimes 24}/\S_{24}$ & 16,610,409,114,771,900
        \\
        $\Dun^{\otimes 24}/\S_{24}^\text{tor}$ & $-$237,043,714,720,252 
        \\
        $\DunArf^{\otimes 24}/\S_{24}^\text{tor}$ & 6,204,518,574,922,375
        \\
        $\Dun^{\otimes 24}/\A_{24}$ & 381,058,359,637,574
        \\
        $\DunArf^{\otimes 24}/\A_{24}$ & 8,306,065,365,519,768
        \\
        \hline
    \end{tabular}
    \caption{The Witten indices of symmetric $\S_n$ and alternating $\A_n$ orbifolds of 24 copies of either the Duncan Supermoonshine module $\Dun$ or its tensor product with the Kitaev chain, $\DunArf := \Dun \otimes (-1)^\text{Arf}$.  The superscript tor denotes discrete torsion. We do not include the results for $\Dun^{\otimes 24}/\A^\text{tor}_{24}$ or   $\DunArf^{\otimes 24}/\A^\text{tor}_{24}$ since these are equivalent to $\DunArf^{\otimes 24}/\A_{24}$ and $\Dun^{\otimes 24}/\A_{24}$ respectively, as discussed in the text. }
    \label{Tab:24}
\end{table}

Though we cannot compute the full Witten indices of the proposed periodicity elements, in the current note we will develop an essential tool for their eventual computation: namely, a closed formula for alternating orbifolds. We present this formula in two forms: one similar to the seminal formula of Dijkgraaf, Moore, Verlinde, Verlinde (DMVV) for symmetric orbifolds \cite{DMVV}, and another involving generalized Hecke operators \cite{baker1990hecke,borcherds1992monstrous,Tuite:2008gy,ganter2009hecke,carnahan2010generalized}.  We also consider orbifolds by subgroups of $\Co_1$ that do not require 
the missing generalized McKay-Thompson data. Having developed this technology, we use it to verify that the Witten indices in the orbifold theories all satisfy the divisibility property demanded by the Stolz-Teichner conjecture.

\subsection{Organization}

The remaining sections are organized as follows.  Section~\ref{Sec:Alternating} presents the second-quantized formula for alternating orbifolds, as well as an expression in terms of generalized Hecke operators. Because the proofs are somewhat long and technical, they are relegated to Appendix \ref{app:Proofs}.  Then in Section~\ref{Sec:Supermoonshine} we use these results to examine divisibility in orbifolds of $\Dun$. In particular, we allow for orbifolds by $\S_n$, $\A_n$, and cyclic subgroups of $\Co_0$ or $\Co_1$, as well as combinations when allowed.  
In addition to the main text, we include two appendices.  Appendix~\ref{Sec:Permutation} contains an analysis of anomalies for permutation symmetries, while Appendix \ref{app:Proofs} contains proofs of statements made in Section \ref{Sec:Alternating}. 

\paragraph{Note:} We thank Theo Johnson-Freyd for explaining the mathematics underlying many of the physical interpretations given in Appendix~\ref{Sec:Permutation}.

\section{Alternating orbifolds}
\label{Sec:Alternating}

As mentioned in the introduction, permutation orbifolds provide a way to construct families of conformal field theories (CFTs) with sparse light spectra. This has been of interest in previous physics literature since the holographic duals in anti-de Sitter space can be weakly coupled \cite{Haehl:2014yla,Keller:2017rtk}.  In contrast, for tensor product theories without any permutation orbifold, the spectrum exhibits Hagedorn growth, since the entropy grows linearly with the central charge.

Symmetric permutation orbifolds have a long history of study, starting most famously with the work of Dijkgraaf, Moore Verlinde, and Verlinde in \cite{DMVV}, to be reviewed below. 
In this section, we present an analogous formula for alternating orbifolds, which in particular gives a closed-form expression for the generating function
\begin{equation}
    \cZ^\A[\T](\sigma,\tau,z) := 2 +  2p\, Z[\T](\tau,z) + \sum_{n=2}^{\infty}p^n Z[\T^{\otimes n}/\A_n](\tau,z)~, \quad p = e^{2\pi i \sigma}~
\end{equation}
of alternating orbifolds for a theory $\cT$. In fact, we will give two closed-form expressions for this quantity, the second involving generalized Hecke operators, similar to a formula of Bantay \cite{Bantay:2000eq} for symmetric orbifolds.  Furthermore, in order to facilitate orbifolding by $\Co_1$ or a subgroup thereof, we will present equivariant formulae that allow for twisting by arbitrary $\Co_1$ elements along both space and time.

Before proceeding, Table~\ref{Tab:Z} lists the different types of torus partition functions that appear, to help the reader navigate our notation.  Following the terminology coined in \cite{Gaberdiel:2012gf}, ``twining'' refers to turning on a flavor fugacity in the trace definition of the torus partition function, which can also be described as ``twisting in the time direction'' or ``inserting a topological defect along the space direction.''

\begin{table}[!t]
    \centering
    \begin{tabular}{|c|c|}
        \hline
        $Z[\T]^g$ & Twined torus partition function of $\T$
        \\
        $\cZ^\text{\Omega}[\T]^g$ & Generating function for twined torus partition functions of $\T^{\otimes n}/\Omega_n$.
        \\
        $\cZ^\text{\Omega}[\T;g]$ & Generating function for torus partition functions of $\T^{\otimes n}/\Omega_n \times \vev{g}$.
        \\\hline
    \end{tabular}
    \caption{Notation for the different torus partition functions appearing in this note, where the permutation group $\Omega$ can be $\S$, $\S^\text{tor}$, $\A$, or $\A^\text{tor}$.}
    \label{Tab:Z}
\end{table}

\subsection{Review of symmetric orbifolds}
\label{Sec:Symmetric}

In \cite{DMVV}, Dijkgraaf, Moore, Verlinde, and Verlinde (DMVV) derived a formula computing the elliptic genera of symmetric orbifolds $\T^{\otimes n}/\S_n$ of a theory $\cT$ in terms of the elliptic genus of $\T$ itself.\footnote{We remind the reader that the elliptic genus considered by DMVV in the $\cN = 2$ context is defined by the following trace over the Ramond sector of the theory,
$$Z[\T](\tau, z)=\Tr_{\Hilb[\T]_\text{R}} (-1)^F q^H y^{J_L}~,$$
where 
$H=L_0-\frac{c}{24}$, $q=e^{2\pi i \tau}$, and $y=e^{2\pi i z}$ is a fugacity for the left-moving $\U(1)_\text{R}$ symmetry.  By contrast, the $\cN = (0,1)$ elliptic genus appearing in the TMF context does not have the $\U(1)$ fugacity.
}
More precisely, the DMVV formula gives a closed-form expression for the generating function
\begin{equation}\label{eq:DMVV}
   \cZ^\S[\cT](\sigma, \tau, z):= 1 + \sum_{n=1}^{\infty} p^n Z[\T^{\otimes n}/\S_n](\tau,z) \end{equation}
of symmetric orbifolds. The formula is as follows,
   \bea
   \label{eq:actualDMVV}
   \cZ^\S[\cT](\sigma, \tau, z)= \prod_{\begin{smallmatrix}n>0 \\m\in \bZ,\ell\end{smallmatrix}} \frac{1}{\left(1-p^n q^m y^\ell\right)}{}_{c(nm,\ell)}~,
\eea
where $c(m,\ell)$ are the coefficients appearing in the expansion of the elliptic genus of $\T$,
\begin{equation}
    Z[\T](\tau,z) = \sum_{m\in\bZ,\ell} c(m,\ell)q^m y^\ell~ 
\end{equation}
and $y = e^{2\pi i z}$ are fugacities for a $\U(1)$ symmetry. 
By restricting to the order $p^n$ terms on both the left and right, this formula allows one to read off an expression for $Z[\T^{\otimes n}/\S_n]$ in terms of the Fourier coefficients of $Z[\cT]$.

Though it will not be important for our purposes in this note, the DMVV formula can be given a physical interpretation in terms of second-quantized strings \cite{DMVV}. Indeed, if we take $\T$ to be a supersymmetric sigma model on a K\"ahler manifold $M$, then each term on the left-hand side of \eqref{eq:DMVV} corresponds to the left-moving partition function of a single string that winds once around the $\S^1$ in a space-time $(M^{\otimes n}/\S_n)\times \S^1\times \mathbb R$. By contrast, the right-hand side realizes the partition function of a second-quantized (left-moving) string in $M\times \S^1$, where the different sectors of momentum $m$, winding $n$, and $F_L=\ell$ have dimensions $|c(nm,\ell)|$. The proof of \eqref{eq:DMVV} exploits the relation between the partition function of a single string with unit winding in $(M^{\otimes n}/\S_n)\times \S^1$ and multiple strings with possibly higher windings in $M\times \S^1$.

For the purposes of this note, it will be useful to have a similar expression for the twisted and twined partition functions of symmetric orbifolds. For twists in the time direction, these are easily incorporated into the DMVV formula. Concretely, say that our starting theory $\cT$ has symmetry $G$. Upon taking the tensor product $\cT^{\otimes n}$, we may consider the diagonal symmetry $G_\text{diag}$ in $G^n$. Given a non-anomalous subgroup $H < G_\text{diag}$ and an element $g \in H$ of order $N$, we may define the twined elliptic genus\footnote{While it was not necessary to include $d$ explicitly since $g^d$ is itself an element of $H$, we choose to do so here to make clear that $Z[\T]^{g^d}(\tau)$ and $c^g(m,\ell)$ are related by a discrete Fourier transform, with $d$ conjugate to $\ell$.  This is in practice how $c^{g^d}(m,\ell)$ can be determined.
}
\ie\label{d}
    Z[\T]^{g^d}(\tau) &= \sum_{m\in\bZ,\ell\in\bZ_{N}} c^g(m,\ell) q^m e^{\frac{2\pi i d \ell}{N}}~,
\fe
where $c^g(m,\ell)$ counts the number of states in the single-copy theory $\T$ with $(L_0 - \frac{c}{24})$-eigenvalue $m$ and $g$-eigenvalue $e^{\frac{2\pi i \ell}{N}}$. 
The generating function for the twined  elliptic genera of the symmetric products is then given by 
\ie\label{twined-DMVV}
    {\cZ^\S[\T]}^{g^d}(\sigma,\tau) &= \prod_{\begin{smallmatrix}n>0 \\m\in\bZ,\ell\in\bZ_{N}\end{smallmatrix}} \frac{1}{(1-p^n q^m e^{\frac{2\pi i \ell}{N}})^{c^g(nm,d\ell)}}~.
\fe

On the other hand, incorporating twists in the spatial direction in this presentation is more difficult. 
Tuite \cite{Tuite:2008gy}, generalizing Bantay \cite{Bantay:2000eq}, provided an alternative expression for $\cZ^\S[\T]$
in terms of generalized Hecke operators \cite{baker1990hecke,borcherds1992monstrous,Tuite:2008gy,ganter2009hecke,carnahan2010generalized}, which allows one to achieve such twists. The starting point is the definition of the $n$-th generalized
Hecke operator $\sfT_n$ acting on a weight-zero modular function, defined as \cite[(15)]{Tuite:2008gy}
\ie \label{eq:Tn}
    \sfT_n Z[\cT]_h^g(\tau) = \frac{1}{n} \sum_{\substack{ad=n \\ 0 \le b < d}} Z[\cT]^{g^a h^b}_{h^d} \left( \frac{a\tau + b}{d} \right)~.
\fe
Here $Z[\T]^g_h$ indicates the Ramond-Ramond torus partition function of $\T$ with a twist $g$ in the temporal direction and another twist $h$ in the spatial direction.\footnote{In terms of topological line operators implementing the symmetry, $h$ and $g$ correspond respectively to lines stretching along the temporal and spatial directions.} Then the generating function  $\cZ^\S[\T]_h^g(\sigma,\tau)$ is given by \cite[(35)]{Tuite:2008gy}
\ie\label{Hecke}
    \cZ^\S[\T]_h^g(\sigma,\tau)
    = \exp\left\{ \sum_{n>0} p^n \sfT_n Z[\T]_h^g(\tau) \right\}~.
\fe
For the case of no spatial twist $h=e$, it can be checked that this formula reduces to the temporally-twisted DMVV formula in (\ref{twined-DMVV}).

Finally, let us mention that in a follow-up of the original work by DMVV, Dijkgraaf introduced a generalization of the DMVV formula for symmetric orbifolds with discrete torsion \cite{Dijkgraaf:1999za}. Since \cite{Schur1911}
\begin{equation}\label{eq:H2SN}
    \H^2(\S_n,\U(1))=\begin{cases} 0 & \text{if}\;\, n<4\\
    \mathbb{Z}_2 & \text{if}\;\, n\geq 4
    \end{cases}
\end{equation}
there is one nontrivial discrete torsion class for symmetric orbifolds with $n\geq 4$, represented by a 2-cocycle $\gamma\in \H^2(\S_n,\U(1))$. Denoting the quotient in the presence of discrete torsion as $\T^{\otimes n}/\S_n^{\text{tor}}$ and defining the generating function 
\ie\cZ^{\S^\text{tor}}[\T](\sigma,\tau,z) :=  1 + \sum_{n=1}^{3} p^n Z[\T^{\otimes n}/\S_n] + \sum_{n=4}^{\infty} p^n Z[\T^{\otimes n}/\S_n^{\text{tor}}]~,
\fe
Dijkgraaf found that\footnote{The corresponding formula in terms of (generalized) Hecke operators was derived by Bantay in \cite{Bantay:2000eq}.}
\begin{align}\label{eq:SNtor}
\cZ^{\S^\text{tor}}[\T]^g 
    &\,=\frac{1}{2}\prod_{\begin{smallmatrix}n>0 \\m\in\bZ,\ell\end{smallmatrix}} \frac{\Bigl( 1 + p^{2n} q^{m + \frac{1}{2}}y^\ell\Bigr)^{c^g (n(2m+1),\ell)} \hspace{-17mm}}{\Bigl( 1 - p^{2n-1} q^{m}y^\ell\Bigr)^{c^g ((2n-1)m,\ell)}\hspace{-17mm}}\hspace{17mm}
    +
    \frac{1}{2}\prod_{\begin{smallmatrix}n>0 \\m\in\bZ,\ell\end{smallmatrix}} \frac{\Bigl( 1 - p^{2n} q^{m + \frac{1}{2}}y^\ell\Bigr)^{c^g (n(2m+1),\ell)} \hspace{-17mm}}{\Bigl( 1 - p^{2n-1} q^{m}y^\ell\Bigr)^{c^g ((2n-1)m,\ell)}\hspace{-17mm}}\hspace{17mm}
    \nonumber \\
    &\:+\frac{1}{2}\prod_{\begin{smallmatrix}n>0 \\m\in\bZ,\ell\end{smallmatrix}} \frac{\Bigl( 1 + p^{2n} q^{m}y^\ell\Bigr)^{c^g (2nm,\ell)} \hspace{-12mm}}{\Bigl( 1 - p^{2n-1} q^{m}y^\ell\Bigr)^{c^g ((2n-1)m,\ell)}\hspace{-17mm}}\hspace{17mm}
    -
    \frac{1}{2}\prod_{\begin{smallmatrix}n>0 \\m\in\bZ,\ell\end{smallmatrix}} \frac{\Bigl( 1 - p^{2n} q^{m}y^\ell\Bigr)^{c^g (2nm,\ell)} \hspace{-12mm}}{\Bigl( 1 - p^{2n-1} q^{m}y^\ell\Bigr)^{c^g ((2n-1)m,\ell)}\hspace{-17mm}}\hspace{17mm}~.
\end{align}

\noindent
Our first goal will be to give analogs of all of these results for alternating orbifolds.

\subsection{Second-quantized formula}\label{sec:AltDMVV}

As discussed in the introduction, our main interest in the current work is in alternating orbifolds. One situation that necessitates alternating orbifolds is when the full permutation group $
\S_n$ is anomalous, and cannot be gauged---see Appendix~\ref{Sec:Permutation} for a discussion of the permutation anomaly.  
In such cases, alternating orbifolds can sometimes still be consistent.  Another circumstance that demands alternating orbifolds, and the one more relevant to this note, is when there is a mixed anomaly between $\S_n$ and another symmetry being gauged. 

We now begin by obtaining a formula analogous to the DMVV formula \eqref{eq:DMVV} for alternating orbifolds $\T^{\otimes n}/\A_n$, i.e.\ orbifolds of $\T^{\otimes n}$ by the subgroup $\A_n\subset \S_n$ of even permutations. We first quote the final result:

\begin{theorem}
\label{thm:AltDMVV}
The generating function for the elliptic genera of alternating orbifolds of a theory $\T$ is given by
\begin{align}\label{eq:AltDMVV}
    \cZ^\A[\T](\sigma,\tau,z) \,&:=  2 +  2p\, Z[\T] + \sum_{n=2}^{\infty}p^n Z[\T^{\otimes n}/\A_n] \\
   &\:\begin{aligned}
   =&\,\frac{1}{2} \prod_{\begin{smallmatrix} n>0\\m\in\bZ,\ell\end{smallmatrix}} \frac{1}{\left(1-p^n q^m y^\ell\right)}{}_{c(nm,\ell)} 
   &&+
   \frac{1}{2} \prod_{\begin{smallmatrix} n>0\\m\in\bZ,\ell\end{smallmatrix}} \frac{1}{\left(1+(-p)^n q^m y^\ell\right)}{}_{c(nm,\ell)}  \nonumber \\
    +&\,
    \frac{1}{2}\prod_{\begin{smallmatrix}n>0 \\m\in\bZ,\ell\end{smallmatrix}} \frac{\Bigl( 1 + p^{2n-1} q^{m}y^\ell\Bigr)^{c((2n-1)m,\ell)} \hspace{-17mm}}{\Bigl( 1 - p^{2n} q^{m+\frac{1}{2}}y^\ell\Bigr)^{c(n(2m+1),\ell)}\hspace{-17mm}}\hspace{12mm}
    &&+
    \frac{1}{2}\prod_{\begin{smallmatrix}n>0 \\m\in\bZ,\ell\end{smallmatrix}} \frac{\Bigl( 1 + p^{2n-1} q^{m}y^\ell\Bigr)^{c((2n-1)m,\ell)} \hspace{-17mm}}{\Bigl( 1 + p^{2n} q^{m+\frac{1}{2}}y^\ell\Bigr)^{c(n(2m+1),\ell)}\hspace{-17mm}}\hspace{17mm}~,
    \end{aligned}
\end{align}
where the coefficients $c(m,\ell)$ are obtained from 
\begin{equation}\label{eq:Z[T]modes}
    Z[\T](\tau,z) = \sum_{m\in\bZ,\ell} c(m,\ell)q^m y^\ell~.
\end{equation}
\end{theorem}

The proof of this formula will be relegated to Appendix \ref{app:altDMVVproof}. Here we will just sketch the general idea. Instead of computing the $\A_n$ orbifold from scratch, we can reuse the results from the $\S_n$ orbifold, keeping only the contributions from even permutations. Concretely, the symmetric orbifold takes the form
\begin{equation}\label{eq:SnOrb}
    Z[\T^{\otimes n}/\S_n]=\frac{1}{|\S_n|}\sum_{\begin{smallmatrix}g,h\in \S_n\\ gh=hg\end{smallmatrix}} Z[\T^{\otimes n}]^g_h~.
\end{equation}
Since the alternating group is already part of this sum, we can get the $\A_n$ orbifold by \textit{projecting out} the contributions from odd permutations,
\begin{align}\label{eq:AnOrb}
    Z[\T^{\otimes n}/\A_n] \,&= \frac{1}{|\A_n|}\sum_{\begin{smallmatrix}g,h\in \A_n\\ gh=hg\end{smallmatrix}} Z[\T^{\otimes n}]^g_h \nonumber\\
    &=\frac{2}{|\S_n|}\sum_{\begin{smallmatrix}g,h\in \S_n\\ gh=hg\end{smallmatrix}} \frac{1}{2}(1+\sgn\, g) \frac{1}{2}(1+\sgn\, h) Z[\T^{\otimes n}]^g_h\nonumber \\
    &=\frac{1}{2}\Big(Z[\T^{\otimes n}/\S_n] + Z[\T^{\otimes n}/\S_n]^\sgn + Z[\T^{\otimes n}/\S_n]_\sgn + Z[\T^{\otimes n}/\S_n]_\sgn^\sgn\Big)~,
\end{align}
where $\mathrm{sgn}(g)$ is the signature of the permutation $g$. 
In the second line, we have inserted the projector $\frac{1}{2}(1+\sgn(\cdot))$ in the $\S_n$ gauging for both the temporal and spatial twists. In the last line, we have repacked the sums in an obvious manner. The final form of \eqref{eq:AnOrb} suggests that this can be interpreted as some sort of $\bZ_2$ gauging and, indeed, this is nothing but the gauging of the $\bZ_2$ subgroup of the quantum symmetry $\text{Rep}(\S_n)$ generated by the representation $\sgn(\cdot)$. As usual, gauging (part of) a quantum symmetry undoes (part of) the original gauging \cite{Vafa:1989ih}. 

We conclude that to compute the alternating orbifold, we may repeat the derivation of the DMVV formula \cite{DMVV} keeping track of the signature of every permutation. That is, the generating function for the elliptic genera of alternating orbifolds is given by
\begin{equation}\label{eq:sgn-gauging}
    \cZ^\A[\T](\sigma,\tau,z)=\frac{1}{2}\left(\mathcal Z_{00} + \mathcal Z_{10} + \mathcal Z_{01} + \mathcal Z_{11}\right)~,
\end{equation}
where $\mathcal Z_{\alpha\beta}$ are the generating functions of the $\S_N$-orbifold elliptic genera with different insertions of $\sgn(\cdot)$ lines, i.e.\
\begin{equation}\label{eq:Zab}
    \mathcal Z_{\alpha\beta} := 1 + \sum_{n=1}^\infty p^n \frac{1}{|\S_n|}\sum_{g,h} (\sgn\, h)^\alpha(\sgn\, g)^\beta Z[\T^{\otimes n}]^g_h~.
\end{equation}
It now only remains to evaluate the quantities $\mathcal Z_{\alpha\beta}$, which is done in Appendix \ref{app:altDMVVproof}.

\subsection{Discrete torsion}\label{sec:discrete-torsion}

Next, we discuss alternating orbifolds with discrete torsion. To do so, let us begin with some more general comments. 
In general, given a 2-cocycle $\gamma$, gauging with discrete torsion on the torus corresponds to weighting each term in the sum \eqref{eq:SnOrb} by a phase \cite{Vafa:1986wx}
\begin{equation}
    \epsilon(g,h)=\frac{\gamma(g,h)}{\gamma(h,g)}~,\qquad gh=hg~.
\end{equation}
There are different ways to interpret this modification. In a path integral formulation, one can interpret this as stacking with a symmetry-protected topological (SPT)  phase before performing the gauging. In the Hamiltonian formulation, in which we are instructed to sum over twisted sectors $\Hilb_h$, the effect of discrete torsion is that instead of keeping the states in $\Hilb_h$ invariant under the centralizer $\cen_h$, we now pick out the states that transform in a non-trivial 1-dimensional representation of $\cen_h$ \cite{Dijkgraaf:1999za}. This representation is given by $\epsilon(\cdot,h)$, which is a homomorphism thanks to the cocycle condition for $\gamma$.

Given two groups $G$ and $A$, a 2-cocycle $\gamma\in \H^2(G,A)$ defines a central extension
\begin{equation}
    1\to A\to \widehat G\to G \to 1
\end{equation}
with multiplication law $\hat g \cdot \hat h=\left(g h,a_g a_h\gamma(g,h)\right)$. This means that one can compute $\epsilon(g,h)$ in terms of the lifted elements $\hat g,\hat h\in \widehat G$ as
\begin{equation}\label{eq:commutator}
    \epsilon(g,h)=[\hat g,\hat h]~.
\end{equation}
Note that although there are in general multiple possible lifts $g,h\to\hat g,\hat h$, the result is independent of this choice since the difference in the lifts lies in the center of $\widehat G$, and thus cancels out in the commutator.

For symmetric orbifolds, the relevant central extension is
\begin{equation}\label{eq:sesShat}
    1\to \mathbb{Z}_2\to \widehat \S_n \to \S_n\to 1\,
\end{equation}
with $\mathbb{Z}_2$ inside $U(1)$. One way to visualize this extension is by embedding it in the lift of $\O(n-1)$ to $\mathrm{Pin}^-(n-1)$ \cite{Dijkgraaf:1999za}. Taking $\S_n$ to be the group of permutations of $n$ orthonormal basis vectors of $\mathbb R^n$, it is clear that $\S_n$ is a finite subgroup of $\O(n-1)$, the symmetry group of the ($n-1$)-dimensional hypersurface connecting the tips of the vectors. Then, the analog of the short exact sequence \eqref{eq:sesShat} is\footnote{There are in fact two different central extensions of $\S_n$ by $\Z_2$; one embedding in $\mathrm{Pin}^-(n-1)$ and the other in $\mathrm{Pin}^+(n-1)$. They are however equivalent when considered as extensions by $U(1)$, so we commit to one of them without loss of generality.}
\begin{equation}
    1\to \mathbb Z_2\to \mathrm{Pin}^-(n-1)\to \O(n-1)\to 1~,
\end{equation}
and the former can actually be completely embedded in the latter.

We now return to the case of alternating orbifolds of interest to us here. To see the possible discrete torsions, we recall that $\H^r(G,\hat A)=\text{Hom}(\H_r(G,A),\U(1))$ (where $\hat A$ denotes the Pontryagin dual of $A$), which equals $\H_r(G,A)$ when these groups are cyclic. We thus have \cite{Schur1911}
\begin{equation}
    \H^2(\A_n,\U(1))=\H_2(\A_n,\mathbb Z)=\begin{cases} 0~, & \text{if}\;\, n<4\\
    \mathbb{Z}_6~, & \text{if}\;\, n=6,7\\
    \mathbb{Z}_2~, & \text{otherwise}~.
    \end{cases}
\end{equation}
For $n\geq 4$ there always exists at least a $\mathbb Z_2$-worth of possible discrete torsions, coming directly from the $\S_n$ case \eqref{eq:H2SN}. This is the only discrete torsion that we will allow for here. It would be interesting to explore the extra possibilities for the cases $n=6,7$. We set $n\geq 4$ for the rest of the discussion.

Since the alternating group $\A_n$ is the subgroup of even permutations in $\S_n$, it corresponds to orientation-preserving transformations when acting on the basis vectors of $\mathbb R^n$, and therefore naturally embeds in $\SO(n-1)$. Then the central extension $\widehat \A_n$ relevant for the $\Z_2$ discrete torsion in alternating orbifolds is realized by the uplift of $\SO(n-1)$ to $\mathrm{Spin}(n-1)$. The upshot is that the discrete torsions for $\A_n$ and $\S_n$ relevant to us are given by the central extensions in the following commutative diagram,
\begin{equation}
    \begin{tikzcd}
        1 \arrow[r] & \Z_2 \arrow[equal]{d} \arrow[r] &\widehat \A_n \arrow[hookrightarrow]{d} \arrow[r] & \A_n \arrow[hookrightarrow]{d} \arrow[r] & 1 \\
        1 \arrow[r] & \Z_2 \arrow[r] &\widehat \S_n \arrow[r] & \S_n \arrow[r] & 1
    \end{tikzcd}
\end{equation}
which can be completely embedded in
\begin{equation}
    \begin{tikzcd}
        1 \arrow[r] &\mathbb Z_2 \arrow[equal]{d} \arrow[r] & \Spin(n-1) \arrow[hookrightarrow]{d} \arrow[r] & \SO(n-1) \arrow[hookrightarrow]{d} \arrow[r] & 1 \\
        1 \arrow[r] &\mathbb Z_2 \arrow[r] & \mathrm{Pin}^-(n-1) \arrow[r] & \O(n-1) \arrow[r] & 1
    \end{tikzcd}
\end{equation}

The above discussion implies that we can again compute the $\A_n$ orbifold with $\mathbb Z_2$ discrete torsion by gauging the quantum symmetry $\sgn(\cdot)$ of $\T^{\otimes n}/\S_n^{\text{tor}}$, i.e. by projecting out the contributions from odd permutations in Dijkgraaf's calculation \cite{Dijkgraaf:1999za}. The final result is then as follows:
\begin{theorem}
\label{thm:alttor}
The generating function $\cZ^{\A^\text{tor}}[\T](\sigma,\tau,z)$ for alternating orbifolds with discrete torsion is given by
\begin{align}
\label{eq:Atorgenfun}
\cZ^{\A^\text{tor}}&[\T](\sigma,\tau,z) := 2 + 2p\, Z[\T] + \sum_{n=2}^{3} p^n Z[\T^{\otimes n}/\A_n] + \sum_{n=4}^{\infty} p^n Z[\T^{\otimes n}/\A_n^{\text{tor}}]\\
    &\:\begin{aligned}
    &\,=\frac{1}{2}\prod_{\begin{smallmatrix}n>0 \\m\in\bZ,\ell\end{smallmatrix}} \frac{\Bigl( 1 + p^{2n} q^{m + \frac{1}{2}}y^\ell\Bigr)^{c(n(2m+1),\ell)} \hspace{-17mm}}{\Bigl( 1 - p^{2n-1} q^{m}y^\ell\Bigr)^{c((2n-1)m,\ell)}\hspace{-17mm}}\hspace{12mm}
    &&+
    \frac{1}{2}\prod_{\begin{smallmatrix}n>0 \\m\in\bZ,\ell\end{smallmatrix}} \frac{\Bigl( 1 - p^{2n} q^{m + \frac{1}{2}}y^\ell\Bigr)^{c(n(2m+1),\ell)} \hspace{-17mm}}{\Bigl( 1 - p^{2n-1} q^{m}y^\ell\Bigr)^{c((2n-1)m,\ell)}\hspace{-17mm}}\hspace{15mm}
    \nonumber \\
    &\,+
    \frac{1}{2} \prod_{\begin{smallmatrix} n>0\\m\in\bZ,\ell\end{smallmatrix}} \left(1+p^n q^m y^\ell\right)^{c(nm,\ell)}
    &&+
    \frac{1}{2} \prod_{\begin{smallmatrix} n>0\\m\in\bZ,\ell\end{smallmatrix}} \left(1-(-p)^n q^m y^\ell\right)^{c(nm,\ell)}~.
    \end{aligned}
\end{align}
\end{theorem}

\noindent
The proof is relegated to Appendix \ref{app:alttor}. Note that this formula differs from \eqref{eq:AltDMVV} only by the replacements $c(m,\ell)\to -c(m,\ell)$, $p\to -p$. This implies that as far as the elliptic genera are concerned, alternating orbifolds with discrete torsion are equivalent (up to a sign) to first stacking $\T$ with $(-1)^{\mathrm{Arf}}$ and then gauging $\A_n$, i.e.~
\begin{equation}\label{eq:equivalence}
    Z[\T^{\otimes n}/\A^{\text{tor}}_n] = (-1)^n Z[\widetilde\T^{\otimes n}/\A_n]~.
\end{equation}

\subsection{Hecke formula}
The formulae \eqref{eq:AltDMVV} and \eqref{eq:Atorgenfun} for alternating orbifolds allow for $\U(1)$ twining, i.e.\ inserting twists in the temporal direction, via the dependence on the fugacity $y$. For $\bZ_N$ instead of $\U(1)$, we may replace $y$ by $e^{\frac{2\pi i}{N}}$ and restrict the range of $\ell$ to $\bZ_N$. These formulae do not, however, allow for twists in the spatial direction that may belong to a different cyclic group $\bZ_M$. For this, we would like formulae in terms of generalized Hecke operators, analogous to that given in (\ref{Hecke}). Defining the generating function for twisted-twined alternating orbifolds, 
\bea
\cZ^\A[\T]^g_h(\sigma, \tau) := 2 +  2p\, Z[\T]^g_h(\tau) +  \sum_{n =2}^\infty p^n Z[\T^{\otimes n}/\A_n]^g_h(\tau)~,
\eea
we have the following result,

\begin{theorem}
\label{thm:genHeckeA}
The generating function $\cZ^\A[\T]^g_h(\sigma, \tau)$ can be written as \bea\label{eq:secQuantAlt}
\cZ^\A[\T]^g_h(\sigma, \tau)= {\frac12} \sum_{\alpha, \beta \in \{0,1\}}  \mathrm{exp}\left\{ \sum_{n >0} p^n \mathsf{T}^{(\alpha, \beta)}_n Z[\T]^g_h (\tau) \right\}~,
 \eea
 with the generalized Hecke operators with characteristics $\mathsf{T}^{(\alpha, \beta)}_n$ defined as
\bea
\label{eq:genHeckeA}
\mathsf{T}^{(\alpha, \beta)}_n Z[\T]^g_h(\tau) := \frac{1}{n} \sum_{\substack{ad=n\\ 0 \leq b < d}}(-1)^{\alpha a(d+1)}(-1)^{\beta((a+1)d+b(d+1))} Z[\T]^{g^a h^b}_{h^d} \left( \frac{a \tau + b}{d}\right)~.
\eea

\end{theorem}

\noindent
If $h=1$ and $g$ is of order $N$, then \eqref{eq:secQuantAlt} reduces to the second-quantized formula with $y = e^{\frac{2\pi i}{N}}$. We relegate the proof of this formula to Appendix \ref{app:Heckeproof}. 

Likewise in the case with discrete torsion, the twisted generating function
\bea
\cZ^{\A^\text{tor}}[\T]^g_h(\sigma, \tau) := 2 +  2p\, Z[\T]^g_h(\tau) +  \sum_{n =2}^{3} p^n Z[\T^{\otimes n}/\A_n]^g_h(\tau)  + \sum_{n=4}^{\infty} p^n Z[\T^{\otimes n}/\A_n^{\text{tor}}]^g_h~\qquad
\eea
is given by the following result,

\begin{corollary}
\label{cor:genHeckeAtor}
The generating function $\cZ^{\A^{\text{tor}}}[\T]^g_h(\sigma, \tau)$ can be written as \bea\label{eq:secQuantAltTor}
\cZ^{\A^{\text{tor}}}[\T]^g_h(\sigma, \tau)= {\frac12} \sum_{\alpha, \beta \in \{0,1\}}  \mathrm{exp}\left\{ \sum_{n >0} p^n (-1)^{n+1} \mathsf{T}^{(\alpha, \beta)}_n Z[\T]^g_h (\tau) \right\}~,
 \eea
where $\mathsf{T}^{(\alpha, \beta)}_n$ are the generalized Hecke operators given in \eqref{eq:genHeckeA}.
\end{corollary}

\noindent 
This simple corollary follows from Theorem \ref{thm:genHeckeA}, recalling that the formulae in Theorems \ref{thm:AltDMVV} and \ref{thm:alttor} are related by the simple replacement $p\to-p$, $Z[\T]\to-Z[\T]$, as noted in \eqref{eq:equivalence}.

\section{Topological modularity of Supermoonshine}
\label{Sec:Supermoonshine}

In the previous section, we reviewed various formulae for symmetric orbifolds of generic theories $\cT$ and introduced analogs for alternating orbifolds. In this section, we apply these formulae to the specific case of the Supermoonshine module $\Dun$ and check that the divisibility property is satisfied. 
We begin with a brief review of the basic features of $\Dun$.

\subsection{Supermoonshine}
\label{Sec:Duncan}

The Supermoonshine module $\Dun$, also known as the Conway SCFT, is a $c=12$ holomorphic SCFT constructed by Duncan \cite{duncan2007super}, which has Conway's largest sporadic simple group $\Co_1$ as a faithful symmetry.  This symmetry is anomalous, realizing the generator of  $\S\H^3(\Co_1) = \H^3(\Co_1, U(1)). \H^2(\Co_1, \ZZ_2) = \ZZ_{12}. \ZZ_2$. The second term corresponds to a projective phase appearing in the R sector only, meaning that $\Co_1$ is realized projectively on $\Duntw$, or equivalently that $\Duntw$ is in a linear representation of the  Schur cover  $\Co_0 = 2.\Co_1$. This theory is one of three self-dual supersymmetric vertex operator algebras (SVOAs) with central charge $c=12$, the others being $V^{fE_8}$ (the theory of 8 chiral bosons based on the $E_8$ root lattice together with their 8 fermionic partners) and $F_{24}$ (the theory of 24 free chiral fermions).

The construction of $\Dun$ is best understood in terms of fermionic extensions of the bosonic vertex operator algebra $V_{\D_{12}}$, as discussed in \cite{Harrison:2018joy}. Here we will briefly review this construction, putting emphasis on the points that will be needed later. First,  $V_{\D_{12}}$ is a VOA of central charge $c=12$ based on the lattice $\D_{12}$, which can be described in terms of representations of the Kac-Moody algebra $\widehat{\mathfrak{so}(24)}_1$. The zero modes of this current algebra generate $\Spin(24)$, and $V_{\D_{12}}$ has four irreducible modules transforming in different representations of this group: the adjoint $A$ (which is $V_{\D_{12}}$ itself), the vector $V$, the spinor $S$, and the conjugate spinor $C$. The last three modules contain only fermionic states, and thus can be used to extend the bosonic $V_{\D_{12}}$ and obtain an SVOA. 

Before discussing the extensions, it is useful to recall the action of $\Spin(24)$ on these representations, in particular the action of its center $\Z_2\times\Z_2$. Let us denote the generator of the first $\Z_2$ by $\eta$ and the generator of the second one by $\Gamma$, such that $\eta$ is in the kernel of the map $\Spin(24)\to \SO(24)$ but $\Gamma\mapsto -\text{Id}$, with $-\text{Id}$ the nontrivial element in the center of $\SO(24)$. The generator $\eta$ acts by $-1$ on the spinor representations, while $\Gamma$ acts as the chirality matrix, i.e.
\begin{alignat}{4}
    &\eta A = A~, \qquad &&\eta V = V~, \qquad &&\eta S = -S~, \qquad &&\eta C = -C~,\nonumber\\
    &\Gamma A = A~, \qquad &&\Gamma V = -V~, \qquad &&\Gamma S = S~, \qquad &&\Gamma C = -C~.
\end{alignat}

With these basic definitions, we are ready to describe the fermionic extensions of $V_{\D_{12}}$.\footnote{By ``extension" here we mean a choice of the content of the NS sector SVOA, with the adjoint $A$ necessarily included. } Extending $A$ by the vector module $V$ yields $F_{24}$ \cite{Harrison:2020wxl}, the theory generated by 24 free fermions $\lambda^i$ transforming in the vector representation of $\SO(24)$. The remaining representations $S$ and $C$ then comprise the R sector, which is called the canonically twisted module $F_{24}^{\text{tw}}$. 
Here the fermion number operator should be such that it flips the sign of all the states in $V$. There are two such elements, namely $\Gamma$ and $\eta \Gamma$. The two choices differ only in their action on the Ramond sector, which is precisely the effect of coupling the theory to the Kitaev chain \cite{Kitaev:2000nmw}, or in continuum language the invertible field theory $(-1)^\text{Arf}$. So we identify the theories with the different choices of $(-1)^F$ as $F_{24}$ and $\widetilde F_{24}:= F_{24}\otimes (-1)^\text{Arf}$. To summarize, we have
\begin{align}
    &F_{24} \cong A\oplus V~, \qquad F_{24}^{\text{tw}}\cong S\oplus C~, \qquad (-1)^F = \Gamma \,\,\,\,\text{ or }\,\,\,\, \eta\Gamma~.
\end{align}
Let us mention in passing that there are several different ways of picking an $\mathcal N=1$ structure on this theory, or in other words different weight-$3/2$ states in $V$ that one can choose as the supercurrent. These supercurrents are linear combinations of cubic terms $\sim\lambda^i\lambda^j\lambda^k$, and they generate 8 different affine Lie algebras of dimension 24. Remarkably, all these structures can be obtained from suitable orbifolds of $V^{fE_8}$, as shown in \cite{Harrison:2020wxl}. 

Next, if $A$ is extended by one of the spinor representations, e.g.\ $S$, we get the Conway SCFT $\Dun$ \cite{duncan2007super}.
In this case, there is only one inequivalent choice of a weight-$3/2$ state in $S$ that can serve as the supercurrent $G(z)$ generating the $\mathcal N=1$ super-Virasoro algebra; all other choices are related by $\Spin(24)$ transformations. Once we choose one such supercurrent, the subgroup of $\Spin(24)$ that leaves $G(z)$ invariant is (isomorphic to) $\Co_0$, the group of automorphisms of the Leech lattice. As we have discussed before, the group $\Co_0$ acts unfaithfully on $\Dun$, and the true symmetry is $\Co_1$. Indeed, the center of this embedding of $\Co_0$ in $\Spin(24)$ coincides with $\{1,\Gamma\}$, which acts trivially on $\Dun$. On the other hand, the twisted sector (a.k.a.\ the R sector) comprises the other two modules $V,C$, on which $\Gamma$ acts by an overall $-1$ sign, indicative of the anomaly in $\Co_1$. 
As before, there are two choices of the fermion number operator, $(-1)^F=\eta\Gamma$ for $\Dun$ and $(-1)^F=\eta$ for $\widetilde{V}^{f\natural}:= \Dun\otimes (-1)^\text{Arf}$. To summarize, we have
\begin{equation}
    \Dun \cong A\oplus S~, \qquad V_{\text{tw}}^{f\natural}\cong V\oplus C~, \qquad (-1)^F = \eta\Gamma\,\,\,\, \text{or }\,\,\,\, \eta~.
\end{equation}

Finally, one can consider the fermionic extension of $V_{\D_{12}}$ by the conjugate spinor module $C$. A priori this is no different from $\Dun$, but if we keep the same choice of weight-$3/2$ state $G(z)\in S$ as before, then we get a new theory $\DunS$ \cite{duncan2015moonshine}.
Since the would-be supercurrent $G(z)$ now lives in the twisted sector of the theory, it cannot be understood as a generator of super-Virasoro symmetry, and hence this theory is not supersymmetric. There are again two choices of fermion number operator due to the Arf invariant, and now $\Co_0$ acts nonfaithfully both in the NS and R sectors.\footnote{Note that $(-1)^F$ is now contained in the center of $\Co_0$ for one of the choices.} To summarize, we have 
\begin{equation}
    \DunS \cong A\oplus C~, \qquad V^{s\natural}_{\text{tw}}\cong V\oplus S~, \qquad (-1)^F = \Gamma\,\,\,\, \text{ or }\,\,\,\, \eta~.
\end{equation}

Note that both modules $\Dun$ and $\DunS$ can be obtained as $\Z_2$ orbifolds of $F_{24}$, as discussed in e.g.\ \cite{Harrison:2020wxl}. Indeed, if we gauge the $\Z_2$ symmetry generated by $(-1)^F=\Gamma$ in $F_{24}$, we first project onto the $\Gamma$-invariant elements of the NS sector $A\oplus V$ and then add the $\Gamma$-invariant elements of the R sector $S\oplus C$. This yields $A\oplus S\cong \Dun$.
Similarly, if we first stack the Kitaev chain on top of $F_{24}$ and then perform the same gauging operation we keep only states invariant under $\eta\Gamma$, resulting in $A\oplus C\cong \DunS$.

\subsection{McKay-Thompson data for Supermoonshine}

Having reviewed the definition of $\Dun$, we would now like to consider various orbifolds of it. Before doing so, it will behoove us to review some properties of the $\Co_0$ symmetry of $\Dun$. 

First, recall that the \emph{McKay-Thompson series} of a holomorphic CFT $\cT$ with global symmetry $G$ refers to the set of torus partition functions $Z[\cT]^g$ twisted along time by elements $g \in G$. They depend only on the conjugacy class $[g]$ of the element $g$.  For a fermionic CFT, there are different McKay-Thompson series for different spin structures.  In the context of TMF, we are mainly interested in the periodic-periodic spin structure, i.e.\ the trace in the Ramond sector with a $(-1)^F$ insertion, which for the theory $\Dun$ is constant due to the $\cN = (1,1)$ supersymmetry.  
In this case, the McKay-Thompson data are given simply by the $\Co_0$ group characters $\chi$ in the 24-dimensional representation; we adopt the convention that
\ie\label{eq:groupcharZ}
    Z[\Dun]^g = - \chi_g~, \qquad Z[\DunArf]^g = \chi_g~.
\fe 
These characters can be found in Table~\ref{Tab:CC}.\footnote{
    Basic group-theoretic data such as character tables are freely available in GAP \cite{GAP4}.  The character tables for $\Co_0$ can be accessed by the command \texttt{tbl:=CharacterTable("2.Co1")}, the class names by \texttt{ClassNames(tbl)}, and the power map by \texttt{List([1..84],x->PowerMap(tbl,x))}.  For $\Co_1$, one replaces \texttt{"2.Co1"} by \texttt{"Co1"}.
}

\begin{table}
    \centering
    \[
        \begin{array}{|c|ccccccccccccccccc|}
            \hline\text{Co}_0 & \text{1A} & \text{2A} & \text{2B} & \text{2C} & \cellcolor{gray!25} \text{4A} & \cellcolor{gray!25} \text{2D} & \text{3A} & \text{6A} & \text{3B} & \text{6B} & \text{3C} & \text{6C} & \cellcolor{gray!25} \text{3D} & \cellcolor{gray!25} \text{6D} & \text{4B} & \text{4C} & \text{4D} \\
            \text{Co}_1 & \text{1A} & \text{1A} & \text{2A} & \text{2A} & \cellcolor{gray!25} \text{2B} & \cellcolor{gray!25} \text{2C} & \text{3A} & \text{3A} & \text{3B} & \text{3B} & \text{3C} & \text{3C} & \cellcolor{gray!25} \text{3D} & \cellcolor{gray!25} \text{3D} & \text{4A} & \text{4A} & \text{4B} \\
            \chi  & 24 & -24 & 8 & -8 & \cellcolor{gray!25} 0 & \cellcolor{gray!25} 0 & -12 & 12 & 6 & -6 & -3 & 3 & \cellcolor{gray!25} 0 & \cellcolor{gray!25} 0 & 8 & -8 & 0 \\
            \hline\text{Co}_0 & \text{4E} & \text{4F} & \cellcolor{gray!25} \text{4G} & \cellcolor{gray!25} \text{8A} & \cellcolor{gray!25} \text{4H} & \text{5A} & \text{10A} & \text{5B} & \text{10B} & \text{5C} & \text{10C} & \text{6E} & \text{6F} & \cellcolor{gray!25} \text{12A} & \text{6G} & \text{6H} & \text{6I} \\
            \text{Co}_1 & \text{4C} & \text{4C} & \cellcolor{gray!25} \text{4D} & \cellcolor{gray!25} \text{4E} & \cellcolor{gray!25} \text{4F} & \text{5A} & \text{5A} & \text{5B} & \text{5B} & \text{5C} & \text{5C} & \text{6A} & \text{6A} & \cellcolor{gray!25} \text{6B} & \text{6C} & \text{6C} & \text{6D} \\
            \chi  & 4 & -4 & \cellcolor{gray!25} 0 & \cellcolor{gray!25} 0 & \cellcolor{gray!25} 0 & -6 & 6 & 4 & -4 & -1 & 1 & -4 & 4 & \cellcolor{gray!25} 0 & -4 & 4 & 5 \\
            \hline\text{Co}_0 & \text{6J} & \text{6K} & \text{6L} & \text{6M} & \text{6N} & \cellcolor{gray!25} \text{6O} & \cellcolor{gray!25} \text{12B} & \cellcolor{gray!25} \text{6P} & \text{7A} & \text{14A} & \text{7B} & \text{14B} & \cellcolor{gray!25} \text{8B} & \cellcolor{gray!25} \text{8C} & \text{8D} & \text{8E} & \text{8F} \\
            \text{Co}_1 & \text{6D} & \text{6E} & \text{6E} & \text{6F} & \text{6F} & \cellcolor{gray!25} \text{6G} & \cellcolor{gray!25} \text{6H} & \cellcolor{gray!25} \text{6I} & \text{7A} & \text{7A} & \text{7B} & \text{7B} & \cellcolor{gray!25} \text{8A} & \cellcolor{gray!25} \text{8B} & \text{8C} & \text{8C} & \text{8D} \\
            \chi  & -5 & 2 & -2 & -1 & 1 & \cellcolor{gray!25} 0 & \cellcolor{gray!25} 0 & \cellcolor{gray!25} 0 & -4 & 4 & 3 & -3 & \cellcolor{gray!25} 0 & \cellcolor{gray!25} 0 & 4 & -4 & 0 \\
            \hline\text{Co}_0 & \text{8G} & \text{8H} & \cellcolor{gray!25} \text{8I} & \text{9A} & \text{18A} & \text{9B} & \text{18B} & \text{9C} & \text{18C} & \text{10D} & \text{10E} & \cellcolor{gray!25} \text{20A} & \cellcolor{gray!25} \text{20B} & \text{10F} & \text{10G} & \text{10H} & \text{10I} \\
            \text{Co}_1 & \text{8E} & \text{8E} & \cellcolor{gray!25} \text{8F} & \text{9A} & \text{9A} & \text{9B} & \text{9B} & \text{9C} & \text{9C} & \text{10A} & \text{10A} & \cellcolor{gray!25} \text{10B} & \cellcolor{gray!25} \text{10C} & \text{10D} & \text{10D} & \text{10E} & \text{10E} \\
            \chi  & 2 & -2 & \cellcolor{gray!25} 0 & -3 & 3 & 0 & 0 & 3 & -3 & -2 & 2 & \cellcolor{gray!25} 0 & \cellcolor{gray!25} 0 & -2 & 2 & 3 & -3 \\
            \hline\text{Co}_0 & \cellcolor{gray!25} \text{10J} & \text{11A} & \text{22A} & \text{12C} & \text{12D} & \text{12E} & \cellcolor{gray!25} \text{12F} & \text{12G} & \text{12H} & \text{12I} & \text{12J} & \cellcolor{gray!25} \text{24A} & \text{12K} & \text{12L} & \text{12M} & \text{12N} & \text{12O} \\
            \text{Co}_1 & \cellcolor{gray!25} \text{10F} & \text{11A} & \text{11A} & \text{12A} & \text{12A} & \text{12B} & \cellcolor{gray!25} \text{12C} & \text{12D} & \text{12D} & \text{12E} & \text{12E} & \cellcolor{gray!25} \text{12F} & \text{12G} & \text{12H} & \text{12H} & \text{12I} & \text{12I} \\
            \chi  & \cellcolor{gray!25} 0 & 2 & -2 & -4 & 4 & 0 & \cellcolor{gray!25} 0 & -1 & 1 & 2 & -2 & \cellcolor{gray!25} 0 & 0 & 1 & -1 & -2 & 2 \\
            \hline\text{Co}_0 & \cellcolor{gray!25} \text{12P} & \text{12Q} & \text{12R} & \cellcolor{gray!25} \text{24B} & \cellcolor{gray!25} \text{12S} & \text{13A} & \text{26A} & \cellcolor{gray!25} \text{28A} & \text{14C} & \text{14D} & \text{15A} & \text{30A} & \text{15B} & \text{30B} & \cellcolor{gray!25} \text{15C} & \cellcolor{gray!25} \text{30C} & \text{15D} \\
            \text{Co}_1 & \cellcolor{gray!25} \text{12J} & \text{12K} & \text{12K} & \cellcolor{gray!25} \text{12L} & \cellcolor{gray!25} \text{12M} & \text{13A} & \text{13A} & \cellcolor{gray!25} \text{14A} & \text{14B} & \text{14B} & \text{15A} & \text{15A} & \text{15B} & \text{15B} & \cellcolor{gray!25} \text{15C} & \cellcolor{gray!25} \text{15C} & \text{15D} \\
            \chi  & \cellcolor{gray!25} 0 & 3 & -3 & \cellcolor{gray!25} 0 & \cellcolor{gray!25} 0 & -2 & 2 & \cellcolor{gray!25} 0 & 1 & -1 & 3 & -3 & -2 & 2 & \cellcolor{gray!25} 0 & \cellcolor{gray!25} 0 & 1 \\
            \hline\text{Co}_0 & \text{30D} & \text{15E} & \text{30E} & \cellcolor{gray!25} \text{16A} & \text{16B} & \text{16C} & \text{18D} & \text{18E} & \text{18F} & \text{18G} & \text{18H} & \text{18I} & \text{20C} & \text{20D} & \cellcolor{gray!25} \text{20E} & \text{20F} & \text{20G} \\
            \text{Co}_1 & \text{15D} & \text{15E} & \text{15E} & \cellcolor{gray!25} \text{16A} & \text{16B} & \text{16B} & \text{18A} & \text{18A} & \text{18B} & \text{18B} & \text{18C} & \text{18C} & \text{20A} & \text{20A} & \cellcolor{gray!25} \text{20B} & \text{20C} & \text{20C} \\
            \chi  & -1 & 2 & -2 & \cellcolor{gray!25} 0 & 2 & -2 & -1 & 1 & 2 & -2 & -1 & 1 & -2 & 2 & \cellcolor{gray!25} 0 & -1 & 1 \\
            \hline\text{Co}_0 & \text{21A} & \text{42A} & \text{21B} & \text{42B} & \cellcolor{gray!25} \text{21C} & \cellcolor{gray!25} \text{42C} & \cellcolor{gray!25} \text{22B} & \cellcolor{gray!25} \text{22C} & \text{23A} & \text{46A} & \text{23B} & \text{46B} & \cellcolor{gray!25} \text{24C} & \text{24D} & \text{24E} & \cellcolor{gray!25} \text{24F} & \cellcolor{gray!25} \text{24G} \\
            \text{Co}_1 & \text{21A} & \text{21A} & \text{21B} & \text{21B} & \cellcolor{gray!25} \text{21C} & \cellcolor{gray!25} \text{21C} & \cellcolor{gray!25} \text{22A} & \cellcolor{gray!25} \text{22A} & \text{23A} & \text{23A} & \text{23B} & \text{23B} & \cellcolor{gray!25} \text{24A} & \text{24B} & \text{24B} & \cellcolor{gray!25} \text{24C} & \cellcolor{gray!25} \text{24D} \\
            \chi  & 2 & -2 & -1 & 1 & \cellcolor{gray!25} 0 & \cellcolor{gray!25} 0 & \cellcolor{gray!25} 0 & \cellcolor{gray!25} 0 & 1 & -1 & 1 & -1 & \cellcolor{gray!25} 0 & -2 & 2 & \cellcolor{gray!25} 0 & \cellcolor{gray!25} 0 \\
            \hline\text{Co}_0 & \cellcolor{gray!25} \text{24H} & \text{24I} & \text{24J} & \cellcolor{gray!25} \text{52A} & \text{28B} & \text{28C} & \cellcolor{gray!25} \text{56A} & \cellcolor{gray!25} \text{56B} & \text{30F} & \text{30G} & \cellcolor{gray!25} \text{60A} & \cellcolor{gray!25} \text{60B} & \text{30H} & \text{30I} & \text{30J} & \text{30K} & \text{33A} \\
            \text{Co}_1 & \cellcolor{gray!25} \text{24E} & \text{24F} & \text{24F} & \cellcolor{gray!25} \text{26A} & \text{28A} & \text{28A} & \cellcolor{gray!25} \text{28B} & \cellcolor{gray!25} \text{28B} & \text{30A} & \text{30A} & \cellcolor{gray!25} \text{30B} & \cellcolor{gray!25} \text{30C} & \text{30D} & \text{30D} & \text{30E} & \text{30E} & \text{33A} \\
            \chi  & \cellcolor{gray!25} 0 & -1 & 1 & \cellcolor{gray!25} 0 & 1 & -1 & \cellcolor{gray!25} 0 & \cellcolor{gray!25} 0 & 1 & -1 & \cellcolor{gray!25} 0 & \cellcolor{gray!25} 0 & 1 & -1 & 0 & 0 & -1 \\
            \hline\text{Co}_0 & \text{66A} & \text{35A} & \text{70A} & \text{36A} & \text{36B} & \text{39A} & \text{78A} & \text{39B} & \text{78B} & \cellcolor{gray!25} \text{40A} & \cellcolor{gray!25} \text{40B} & \cellcolor{gray!25} \text{84A} & \text{60C} & \text{60D} & & & \\
            \text{Co}_1 & \text{33A} & \text{35A} & \text{35A} & \text{36A} & \text{36A} & \text{39A} & \text{39A} & \text{39B} & \text{39B} & \cellcolor{gray!25} \text{40A} & \cellcolor{gray!25} \text{40A} & \cellcolor{gray!25} \text{42A} & \text{60A} & \text{60A} & & & \\
            \chi  & 1 & 1 & -1 & -1 & 1 & 1 & -1 & 1 & -1 & \cellcolor{gray!25} 0 & \cellcolor{gray!25} 0 & \cellcolor{gray!25} 0 & 1 & -1 & & & \\\hline
        \end{array}
    \]
    \caption{The 167 conjugacy classes of $\Co_0$, their projections to the 101 conjugacy classes of $\Co_1$, and the $\Co_0$ group characters $\chi$ in the 24-dimensional irreducible representation.  The anomalous classes in Supermoonshine $\Dun$ are shaded.}
    \label{Tab:CC}
\end{table}

Let us now discuss the Conway anomaly.  The Conway group $\Co_0$ was discovered in \cite{conway1968perfect,conway1969group} and is known to have 167 conjugacy classes,
43 of them are anomalous \cite[Theorem~7.1]{Johnson-Freyd:2017bpn}:
\ie\label{AnomalousClasses}
    & \text{4A, 2D, 3D, 6D, 4G, 8A, 4H, 12A, 6O, 12B, 6P, 8B, 8C, 8I, 20A, 20B,}
    \\
    & \text{10J, 12F, 24A, 12P, 24B, 12S, 28A, 15C, 30C, 16A, 20E, 21C, 42C, 22B,}
    \\
    & \text{22C, 24C, 24F, 24G, 24H, 52A, 56A, 56B, 60A, 60B, 40A, 40B, 84A.}
\fe 
As for $\Co_1$, the anomalous classes are given by the pullback $\S\H^3(\Co_1) \to \S\H^3(\Co_0)$ which in this case is an isomorphism; among the 101 classes, 37 are anomalous in $\Dun$:
\ie\label{AnomalousClassesCo1}
    & \text{2B, 2C, 3D, 4D, 4E, 4F, 6B, 6G, 6H, 6I, 8A, 8B, 8F, 10B, 10C,}
    \\
    & \text{10F, 12C, 12F, 12J, 12L, 12M, 14A, 15C, 16A, 20B, 21C,}
    \\
    & \text{22A, 24A, 24C, 24D, 24E, 26A, 28B, 30B, 30C, 40A, 42A.}
\fe 
Note that the anomaly forces the twined Witten indices of these classes to vanish.
The data on anomalies is again collected in Table~\ref{Tab:CC}.

When we consider orbifolds by cyclic subgroups of $\Co_0$ or $\Co_1$, only those outside of these anomalous classes make physical sense.
However, because $\Dun^{\otimes n}/\A_n$ has $\Co_1$ symmetry only when $n$ is even, for the purpose of Section~\ref{Sec:AltSuper} let us also record the 23 conjugacy classes in $\Co_1$ that are anomalous in $\Dun^{\otimes n}$ for $n$ even:
\ie\label{DoubleAnomalousClassesCo1}
    & \text{2B, 3D, 4E, 4F, 6B, 6H, 6I, 8F, 10B, 10C, 12F, 12L, 12M,}
    \\
    & \text{14A, 15C, 20B, 21C, 24D, 26A, 28B, 30B, 30C, 42A.}
\fe

Finally, a technical property of $\Dun$ that simplifies the orbifold computations is the following: for every $g \in \Co_0$, 
\ie\label{PowerMapProperty}
    \eg[\T]^{g^r} = \eg[\T]^{g^{\mygcd{r,N}}} \quad
    \forall\,\, r,
\fe
where $N$ is the order of $g$.

\subsection{Symmetric orbifolds of Supermoonshine}

We may now proceed to orbifolds of Supermoonshine, starting with symmetric orbifolds. Consider the second-quantized formula \eqref{eq:DMVV} for symmetric orbifolds, but with a $\bZ_N = \vev{g}$ symmetry instead of $\U(1)$.  In the presence of $\cN=(1,1)$ supersymmetry, the elliptic genera are constants given by the twined Witten indices; in other words, the Fourier expansion in (\ref{d}) becomes simply
\bea
Z[\Dun]^{g^k}(\tau) = \sum_{\ell \in \ZZ_N} c^g(0,\ell) e^{2\pi i \ell k \over N}~.
\eea
From now on we will write $c^g(\ell) := c^g(0,\ell)$.
Due to \eqref{PowerMapProperty}, $Z[\Dun]^{g^k}(\tau)$ only depends on $\gcd(k,N)$.  The second-quantized formula \eqref{eq:DMVV} becomes
\ie
    \EG^\text{S}[\Dun]^g(\sigma) 
    &= \prod_{\substack{n>0 \\ \ell\in\bZ_N}} \frac{1}{(1-p^ny^\ell)^{c^g(\ell)}}
    = \prod_{\ell\in\bZ_N} \frac{1}{(py^\ell;p)^{c^g(\ell)}}~,
\fe
where $y = e^{2\pi i/N}$ and $(a;q) = (a;q)_\infty$ is the $q$-Pochhammer symbol.
For $g=e$, if we let $\wi = \eg[\Dun]$ be shorthand for the (untwisted) Witten index, then the generating function for the symmetric orbifold Witten indices simplifies to
\ie\label{SymmetricOrbifoldGeneratingFunction}
    \EG^\text{S}[\Dun](\sigma) &= \frac{1}{(p;p)^\wi} = \frac{p^{\frac{\wi}{24}}}{\eta(\sigma)^\wi}~.
\fe
  For Supermoonshine $\Dun$ the Witten index is $\cI = -24$, while for $\widetilde{V}^{f \natural}:=\Dun \otimes (-1)^{\mathrm{Arf}} $ the Witten index is $\cI = 24$. The elliptic genus for  $\Dun^{\otimes n}/\S_{n}$ and $(\widetilde{V}^{f \natural})^{\otimes n}/\S_{n}$ can then be obtained by Fourier expanding (\ref{SymmetricOrbifoldGeneratingFunction}) and 
reading off the order $p^n$ Fourier coefficient, e.g.
\bea
Z[\Dun^{\otimes 2}/\S_{2}] &=& 252~, \hspace{0.66 in}Z[(\widetilde{V}^{f \natural})^{\otimes 2}/\S_{2}] = 324 ~.
\no\\
Z[\Dun^{\otimes 3}/\S_{3}] &=& -1472 ~,\hspace{0.46 in}Z[(\widetilde{V}^{f \natural})^{\otimes 3}/\S_{3}] =3200 ~.
\no\\
Z[\Dun^{\otimes 4}/\S_{4}] &=&4830 ~,\hspace{0.58 in}Z[(\widetilde{V}^{f \natural})^{\otimes 4}/\S_{4}] =25650 ~.
\eea
In particular, the Witten index of $\Dun^{\otimes 24}/\S_{24}$ is $-$25,499,225, while that of $\DunArf^{\otimes 24}/\S_{24}$ is 16,610,409,114,771,900. These are collected in Table \ref{Tab:24}. In all of the cases listed above, we see that divisibility is satisfied---namely that $Z[\Dun^{\otimes n}/\S_{n}]$ and $Z[(\widetilde{V}^{f \natural})^{\otimes n}/\S_{n}]$  are divisible by ${24 / \mathrm{gcd}(24,n)}$.

In fact, we may prove this divisibility for arbitrary $n$ by using the fact given in (\ref{Divisibility}) and explained in footnote \ref{footnote:1}, namely that proving divisibility of  $\cZ^\S[\cT]|_{p^n}$ by ${24 / \mathrm{gcd}(24,n)}$ is equivalent to proving divisibility of $n \cZ^\S[\cT]|_{p^n}$ by $24$. In other words, it suffices to show that
\bea
\label{eq:easierdivisibility}
24 \,\, \Big|\,\,p { d \cZ^\S[V^{f \natural}] \over dp}~. 
\eea
 To this effect, we define $\phi(p) = p^{1\over 24} / \eta(p)$ such that $\cZ^\S[V^{f \natural}](\sigma) = \phi(p)^{\pm 24}$, whence
\ie  
    \frac{d\cZ^\S[V^{f \natural}](\sigma)}{dp} =   \frac{d\phi(p)^{\pm 24}}{dp} = \pm 24 \, \phi(p)^{\pm24-1} \frac{d\phi(p)}{dp}~.
\fe
Since $\phi(p)$ has integer Fourier coefficients, the divisibility property \eqref{eq:easierdivisibility} is automatically satisfied.

\paragraph{Orbifolds by cyclic subgroups of Co$_\text{0}$}  Due to \eqref{PowerMapProperty}, if $g$ is non-anomalous then 
the orbifold Witten indices can be written as
\ie\label{GenOrb}
   Z[\cT/\langle g \rangle](\tau) = \sum_{d \dv N} \frac{J_2(N/d)}{N} Z[\cT]^{g^d}(\tau)~,
\fe
where $J_2(N)$ is the Jordan totient function of $N$, i.e. the number of pairs $(m,n)$ such that $m,n\leq N$ and $\mathrm{gcd}(m,n,N)=1$. It admits the following closed-form expression, 
\bea
J_2(N) = N^2 \prod_{p|N}\left(1 - {1\over p^2} \right)~.
\eea
 
Using the formula above we may now compute the orbifold Witten indices for $\cT/\langle g \rangle$ given the twisted indices $Z[\cT]^{g^d}$.  Fortunately in the case of $\cT = \Dun$, the latter are simply given by the characters of the 24-dimensional representation of $\Co_0$, as in (\ref{eq:groupcharZ}).
For example, if $g_{2\text{A}} \in 2 \text{A}$ and $g_{3\text{A}} \in 3 \text{A}$ one has 
\bea
Z[\Dun] &=& -24~, \hspace{0.38 in}
Z[\Dun]^{g_{2\mathrm{A}}} = 24~, \hspace{0.38 in}
Z[\Dun]^{g_{3\mathrm{A}}} = Z[\Dun]^{g_{3\mathrm{A}}^2} = 12~, 
\eea
in the usual notation for the conjugacy classes of $\Co_0$. From these one can then compute 
\bea
Z[\Dun/\langle 2\mathrm{A}\rangle] = 24~, 
\hspace{0.5 in}
Z[\Dun/\langle 3\mathrm{A}\rangle] = 24~. 
\eea
We note that both of the above are divisible by 24, consistent with the divisibility property (\ref{Divisibility}). Indeed, computer implementation allows one to check divisibility for all 167 conjugacy classes of $\Co_0$. For some values of $n$ divisibility is actually found to be violated, but this occurs only when $g$ belongs to the 43 anomalous conjugacy classes given in \eqref{AnomalousClasses}, and hence no such gauging was allowed in the first place; in particular, for $n=1$, divisibility is violated precisely for those 43 classes. Note that for certain values of $n$, the conjugacy classes in \eqref{AnomalousClasses} can actually become non-anomalous, and in those cases we again find that $\Dun^{\otimes n}/\S_n \times \langle g \rangle$ satisfies divisibility.

We may furthermore compute $Z[\cT^{\otimes n}/\S_n \times \langle g \rangle]$ by combining (\ref{GenOrb}) with (\ref{twined-DMVV}), 
\bea
\cZ^\S[\cT; g](\sigma) := \sum_{d \dv N} \frac{J_2(N/d)}{N} \prod_{\begin{smallmatrix}n>0 \\m\in\ZZ,\ell\in\ZZ_{N}\end{smallmatrix}} {1\over (1-p^n q^m e^{2 \pi i \ell \over N})^{c^{g^d}(nm, \ell)}}~
\eea
where $c^g(m, \ell)$ are the Fourier coefficients of $Z[\cT]^{g^d}(\tau)$, as in (\ref{d}). For $\cT = \Dun$, we find 
\bea
 \cZ^\S[\Dun;  2\mathrm{A}](\sigma) &=& - 1 + {3\over 2} (-p\,;\,p)^{24} + \half { (p\,;\,p)^{24}} 
 \no\\
 &=& 1 + 24\,p + 576\,p^2 + 3200\,p^3 + 29604\,p^4 + 155232\,p^ 5 + \dots
\no\\\no\\
 \cZ^\S[\Dun;  3\mathrm{A}](\sigma) &=& -2 + {1\over 3} { (p\,;\,p)^{24}} + {8\over 3} (e^{2\pi i \over 3} p \,;\, p)^{12} (e^{-2\pi i \over 3} p \,;\, p)^{12}
 \no\\
 &=& 1 + 24\,p + 324\,p^2 + 864\,p^3 + 7986\,p^4 + 24192\,p^5 + \dots
\eea
and so on.  The indices $Z[\Dun^{\otimes n}/\S_n \times \langle g \rangle]$ are then obtained by taking the order $p^n$ terms in the Fourier expansions of the generating functions, e.g.\ 
\ie 
    Z[\Dun^{\otimes n}/\S_n \times \langle g_{2\text{A}} \rangle] &= 576,\, 3200,\, 29604, \hspace{0.34 in} n = 2,3,4,
    \\
    Z[\Dun^{\otimes n}/\S_n \times \langle g_{3\text{A}} \rangle] &= 324,\, 864,\, 7986, \hspace{0.5 in} n = 2,3,4.
\fe
One may check that each of these is divisible by $24 / \mathrm{gcd}(24, n)$, as required by the divisibility constraint. Computer implementation allows one to check this for all 167 conjugacy classes of $\Co_0$ and verifies that $(\Dun)^{\otimes n}/\S_n \times \vev{g}$ satisfies divisibility for all $n$ for every conjugacy class outside of \eqref{AnomalousClasses}.

\paragraph{Discrete torsion}

Let us finally mention the case of symmetric orbifolds with discrete torsion. The expression for this was given in (\ref{eq:SNtor}); taking $\cT = V^{f \natural}$, it reduces to
\ie\label{ZgSymTorsion}
    \EG^{\S^\text{tor}}[\Dun]^g(\sigma) = \frac12 \prod_{\ell\in\bZ_N} \frac{(-p^2y^\ell;p^2)^{c^g(\ell)}}{(py^\ell;p^2)^{c^g(\ell)}} - \frac12 \prod_{\ell\in\bZ_N} \frac{(p^2y^\ell;p^2)^{c^g(\ell)}}{(py^\ell;p^2)^{c^g(\ell)}} + \prod_{\ell\in\bZ_N} \frac{1}{(py^\ell;p^2)^{c^g(\ell)}}~.
\fe
For $g=e$ we have 
\bea
\label{eq:Storgenfunc}
 \EG^{\S^\text{tor}}[\Dun](\sigma) &=& \half {(p;p^2)^{24}  \over (-p^2; p^2)^{24}} - \half {(p;p^2)^{24} \over (p^2; p^2)^{24} } + {(p;p^2)^{24}} 
 \no\\
 &=& 1-24 p + 252 p^2 -1472 p^3 + 4554 p^4 + 576 p^5 + \dots
\eea
For $n\geq 4$, we can read off $Z[\Dun^{\otimes n}/ \S_n^\text{tor}]$ from the order $p^n$ term in this expansion, while for $n<4$ we have $Z[\Dun^{\otimes n}/ \S_n^\text{tor}] = Z[\Dun^{\otimes n}/ \S_n]$. Thus we find
\bea
Z[\Dun^{\otimes 2}/ \S_2^\text{tor}] &=& 252~, \hspace{0.5 in} Z[\Dun^{\otimes 3}/ \S_3^\text{tor}] = -1472~, 
\no\\
Z[\Dun^{\otimes 4}/ \S_4^\text{tor}] &=& 4554~, \hspace{0.44 in} Z[\Dun^{\otimes 5}/ \S_5^\text{tor}] = 576~, 
\eea
and so on. In particular, the Witten index of $\Dun^{\otimes 24}/\S_{24}^\text{tor}$ is $-$237,043,714,720,252, which is recorded in Table \ref{Tab:24}. We see that these values are always divisible by $24 / \mathrm{gcd}(24, n)$, as required by the divisibility constraint.  Similar statements hold for $\DunArf$.

We can also consider orbifolds by cyclic subgroups of $\Co_0$ by combining (\ref{GenOrb}) with (\ref{eq:SNtor}). The closed-form expressions in terms of Pochammer symbols are straightforwardly obtained but messy, so here we only record a couple of explicit examples, 
\bea
\cZ^{\S^\text{tor}}[\Dun; {2\text{A}}](\sigma) &=& 1 + 24 p + 576 p^2 + 3200 p^3 + 29052 p^4 + 148608 p^5 + \dots
\no\\
\cZ^{\S^\text{tor}}[\Dun; {3\text{A}}](\sigma) &=& 1 + 24 p + 324 p^2 + 864 p^3 + 7686 p^4 + 23904 p^5 + \dots
\eea
from which we can read off e.g.
\bea
 Z[(\Dun)^{\otimes 4}/ \S_4^\text{tor} \times \langle g_{2 \text{A}} \rangle] &=& 29052~, \hspace{0.38 in}  Z[(\Dun)^{\otimes 5}/ \S_5^\text{tor} \times \langle g_{2 \text{A}} \rangle] = 148608~, 
\no\\
 Z[(\Dun)^{\otimes 4}/ \S_4^\text{tor} \times \langle g_{3 \text{A}} \rangle] &=& 7686~, \hspace{0.45 in} Z[(\Dun)^{\otimes 5}/ \S_5^\text{tor} \times \langle g_{3 \text{A}} \rangle] = 23904~. 
\eea
These can again be checked to satisfy the divisibility criterion. Computer implementation allows us to check this for all conjugacy classes of $\Co_0$, with the list of conjugacy classes violating divisibility being the same as that in the absence of discrete torsion.

\subsection{Alternating orbifolds of Supermoonshine}
\label{Sec:AltSuper}

Next consider the second-quantized formula \eqref{eq:AltDMVV} for alternating orbifolds, but with a $\bZ_N = \vev{g}$ symmetry instead of $\U(1)$.  
In the presence of $\cN=(1,1)$ supersymmetry, we have
\ie\label{ZgAlt}
    & \EG^\text{A}[\Dun]^g(\sigma) 
    \\
    &= \frac12 \prod_{\substack{n>0 \\ \ell\in\bZ_N}} \frac{1}{(1-p^ny^\ell)^{c^g(\ell)}} + \frac12 \prod_{\substack{n>0 \\ \ell\in\bZ_N}} \frac{1}{(1+(-p)^ny^\ell)^{c^g(\ell)}} + \prod_{\substack{n>0 \\ \ell\in\bZ_N}} (1+p^{2n-1}y^\ell)^{c^g(\ell)}
    \\
    &= \frac12 \prod_{\ell\in\bZ_N} \frac{1}{(py^\ell;p)^{c^g(\ell)}} + \frac12 \prod_{\ell\in\bZ_N} \frac{1}{(py^\ell;-p)^{c^g(\ell)}} + \prod_{\ell\in\bZ_N} {(-py^\ell;p^2)^{c^g(\ell)}}~,
\fe
expressed in terms of quantities defined in the previous subsection.
For $g=e$, we have
\ie\label{AlternatingOrbifoldGeneratingFunctionNoRewrite}
    \EG^\text{A}[\Dun](\sigma) &= \frac12 \frac{1}{(p;p)^\wi} + \frac12 \frac{1}{(p;-p)^\wi} + {(-p;p^2)^\wi}~,
\fe
which admits a rewriting as
\ie\label{AlternatingOrbifoldGeneratingFunction}
    \EG^\text{A}[\Dun](\sigma) &=
    \frac12 \frac{1}{(p;p)^\wi} + \frac32 \frac{1}{(p;-p)^\wi}
    = \frac12 \frac{p^{\frac{\wi}{24}}}{\eta(\sigma)^\wi} + \frac32 \frac{p^{\frac{\wi}{24}} \eta(2\sigma)^{2\wi}}{\eta(\sigma)^\wi \eta(4\sigma)^\wi}~.
\fe
The elliptic genus for  $\Dun^{\otimes n}/\A_{n}$ and $(\widetilde{V}^{f \natural})^{\otimes n}/\A_{n}$ can then be obtained by Fourier expanding (\ref{AlternatingOrbifoldGeneratingFunction}) and 
reading off the order $p^n$ Fourier coefficient, e.g.
\bea
Z[\Dun^{\otimes 2}/\A_{2}] &=& 576~, \hspace{0.765 in}Z[(\widetilde{V}^{f \natural})^{\otimes 2}/\A_{2}] = 576~,
\no\\
Z[\Dun^{\otimes 3}/\A_{3}] &=& -4672 ~,\hspace{0.57 in}Z[(\widetilde{V}^{f \natural})^{\otimes 3}/\A_{3}] = 4672 ~,
\no\\
Z[\Dun^{\otimes 4}/\A_{4}] &=& 29604 ~,\hspace{0.625 in}Z[(\widetilde{V}^{f \natural})^{\otimes 4}/\A_{4}] = 29628 ~.
\eea
In particular, the Witten index of $\Dun^{\otimes 24}/\A_{24}$ is 381,058,359,637,574, while that of $\DunArf^{\otimes 24}/\A_{24}$ is 8,306,065,365,519,768. These are collected in Table \ref{Tab:24}. In all of the cases listed above, we see that divisibility is satisfied---namely that $Z[\Dun^{\otimes n}/\A_{n}]$ and $Z[(\widetilde{V}^{f \natural})^{\otimes n}/\A_{n}]$  are divisible by ${24 / \mathrm{gcd}(24,n)}$. 

Incidentally, note that $\cZ^{\A}[V^{f\natural}](\sigma)$ can be written in terms of the McKay-Thompson series $T_{4\A}(\sigma)$ for the $4\A$ conjugacy class of the Monster, 
\bea
\cZ^{\A}[V^{f\natural}](\sigma)= {p \over 2} \left[\Delta(\sigma)^{-1} + 3 T_{4\A}(\sigma) \right] ~. 
\eea
If we define the Fourier coefficients of the inverse modular discriminant and $T_{4\A}(\sigma)$ as follows, 
\bea
\Delta(\sigma)^{-1} = p^{-1} \sum_{n=1}^\infty \overline{\tau}_n \,p^n~, \hspace{0.5in} T_{4\A}(\sigma) = p^{-1} \sum_{n=1}^\infty c^{4\A}_n \,p^n~,
\eea
then we have
\bea
\cZ^{\A}[V^{f\natural}](\sigma)= \sum_{n=0}^{\infty}\half\left(\overline{\tau}_n + 3 c^{4\A}_n\right) p^n~
\eea
and the divisibility conjecture becomes the statement that
\bea
\label{eq:assumption}
{24 \over \mathrm{gcd}(24,n)} \,\, \Big |\,\, \half \left(\overline{\tau}_n + 3 c^{4\A}_n\right)~. 
\eea
Unfortunately, we will not be able to prove this result, but we have verified it to extremely large values in $n$. 

Note that proving this would also prove the divisibility for symmetric orbifolds with discrete torsion, since (\ref{eq:Storgenfunc}) can be rewritten as 
\bea
\cZ^{\S^{\text{tor}}}[V^{f \natural}](\sigma)= {p^3 \over 2} (T_{2\B}(\sigma)-24)^2\left[T_{4\A}(\sigma)-\Delta(\sigma)^{-1} \right]~.
\eea
For the purposes of divisibility we may drop the factor of 24 above, and then defining the Fourier expansion of  $T_{2\B}(p)$ to be 
\bea
 T_{2\B}(\sigma) = p^{-1}\sum_{n=0}c^{2\B}_n\,p^n~,
\eea
we find that 
\bea
{p^3 \over 2} T_{2\B}(\sigma)^2\left[T_{4\A}(\sigma)-\Delta(\sigma)^{-1} \right]=\sum_{n=0}^\infty \sum_{m=0}^n {1 \over2} c^{2\B}_{n-m}(c^{4\A}_m-\overline{\tau}_m)p^n~.
\eea
Proving divisibility thus amounts to proving that 
\bea
\label{eq:symprodtor1}
{24 \over \mathrm{gcd}(24,n)}\,\,\Big |\,\, {1\over 2}\sum_{m=0}^nc^{2\B}_{n-m}(c^{4\A}_m-\overline{\tau}_m)~.
\eea
Assuming (\ref{eq:assumption}), there exists an integer $k_m$ such that 
\bea
\label{eq:importantsimplification}
\half \overline{\tau}_m = {24 \over \mathrm{gcd}(24,m)}k_m - {3 \over 2} c^{4\A}_m~, 
\eea
and plugging this into the right of (\ref{eq:symprodtor1})  gives 
\bea
\sum_{m=0}^n c^{2\B}_{n-m}\left( 2 c^{4\A}_m - {24 \over \mathrm{gcd}(24,m)}k_m \right) ~.
\eea
We then need only prove that 
\bea
\label{eq:divisibilityMT}
{24\over \mathrm{gcd}(24,n)}\,\,\Big|\,\, c^{2\B}_n \hspace{0.4 in}\text{and}\hspace{0.4 in}{24\over \mathrm{gcd}(24,n)}\,\,\Big|\,\, c^{4\A}_n~, 
\eea
since this together with the fact that 
\bea
\label{eq:combinegcds}
{24\over \mathrm{gcd}(24,n)} \,\,\Big|\,\,{24\over \mathrm{gcd}(24,n-m)} \cdot{24\over \mathrm{gcd}(24,m)}
\eea
would imply  (\ref{eq:symprodtor1}). 

Both of the claims in (\ref{eq:divisibilityMT}) are shown in a similar manner by taking a derivative and verifying (\ref{eq:easierdivisibility}). For example, the $T_{4\A}(p)$ McKay-Thompson series
\bea
T_{4\A}(p) = {\Delta(p^2)^2 \over \Delta(p) \Delta(p^4)} = \left(\eta(p^2)^2 \over \eta(p) \eta(p^4) \right)^{24}~
\eea
is clearly the 24th power of a modular form with integer Fourier coefficients, and hence taking a derivative gives a function whose Fourier coefficients are all divisible by 24. Similar comments hold for  $T_{2\B}(p)$.

\paragraph{Unfaithfulness of Co$_\text{0}$}
Returning to alternating orbifolds, we now argue that the central element $z \in \Co_0$ is not faithful in $\T^{\otimes n}/\A_n$ for $n$ even, at the level of Witten indices.  
Since $\ch(e) = \wi$ and $\ch(z) = -\wi$, we have $c^z(0) = 0$ and $ c^z(1) = \wi$ for $\wi \ge 0$, as well as  $c^z(0) = -\wi$ and  $c^z(1) = 0$ for $\wi \le 0$.  In the former case, only $\ell = 1$ contributes to each term in \eqref{ZgAlt}, giving
\ie\label{Unfaithful}
    \EG^\text{A}[\Dun]^z(\sigma) = \frac12 \frac{1}{(-p;p)^\wi} + \frac12 \frac{1}{(-p;-p)^\wi} + {(p;p^2)^\wi}.
\fe
Comparing with  \eqref{AlternatingOrbifoldGeneratingFunctionNoRewrite}, the untwined 
\ie
    2 \EG^\text{A}[\Dun](\sigma)|_{p^\text{even}} = \EG^\text{A}[\Dun](\sigma) + \EG^\text{A}[\Dun](\sigma+1/2)
\fe
and twined
\ie
    2 \EG^\text{A}[\Dun]^z(\sigma)|_{p^\text{even}} = \EG^\text{A}[\Dun]^z(\sigma) + \EG^\text{A}[\Dun]^z(\sigma+1/2)
\fe
generating functions for $n$ even are manifestly equal, suggesting that $\Co_0$ is not faithful, and hence that the $\Co_1$ symmetry is non-anomalous.  The case of $\cI \le 0$ proceeds in the same way.

\paragraph{Orbifolds by cyclic subgroups of Co$_{\text{1}}$}  
We may now consider orbifoldings by subgroups of $\Co_1$. If $g$ is non-anomalous, the orbifold Witten indices take the same form as in \eqref{GenOrb}, with $Z[\cT]^{g^d}(\tau)$ replaced by the generating function for alternating orbifolds.
We will not bother to write the formulae out explicitly in this case, but simply note the result that the would-be orbifold Witten indices of $(\DunArf)^{\otimes n}/\A_n \times \vev{g}$ satisfy divisibility for \textit{all even $n$ and for every $g \in \Co_1$} (to reemphasize, we restrict to $n$ even since only then do we have a non-anomalous $\Co_1$ symmetry). Interestingly, the divisibility in this case holds regardless of any anomalies, echoing the observations for Monstrous Moonshine (Monster CFT) made in \cite{Lin:2022wpx}.  On the other hand, the would-be orbifold Witten indices of $(\Dun)^{\otimes n}/\A_n \times \vev{g}$ do violate divisibility for some even values of $n$, but this occurs only when $g$ belongs to these 8 conjugacy classes of $\Co_1$:
\ie\label{ClassesViolatingDivisibilityArf}
    \text{3D, 6H, 6I, 12L, 12M, 15C, 21C, 30C,}
\fe
all of which are anomalous, c.f. the list \eqref{DoubleAnomalousClassesCo1}.
Note that for some $n$ the anomalies of these conjugacy classes in the diagonal $\Co_1$ are trivialized, and in those cases divisibility is again satisfied.

\subsection{Saturation of divisibility by decomposable theories}

Finally, as an aside, let us try to construct decomposable theories (i.e.\ theories with multiple vacua  at finite volume) saturating the divisibility criterion.
In other words, we search for theories with $\nu = 2(c_\text{R} - c_\text{L}) = -24n$ and Witten index $24/\gcd(24,n)$ for any $n$.  Since we can take direct sums, the question of constructability amounts to whether the greatest common divisor of the Witten indices of a pair, or more generally a tuple, of theories is equal to $24/\gcd(24,n)$.  It turns out that it suffices to consider symmetric orbifolds without discrete torsion for $\Dun$ and $\DunArf$, further gauged by non-anomalous cyclic subgroups of $\Co_0$  (in particular, alternating orbifolds are not necessary, though they diversify the possible constructions).  Let us denote the gcd for these hundred or so theories by $\mathfrak{I}_n$. These quantities may be obtained via computer evaluation and are collected in Table~\ref{Tab:SaturationNaive} for $n = 0, 1, \dotsc, 24$.  We see that for $n \neq 6, 9, 14, 22$ mod $24$, the gcd is precisely $24 / \mathrm{gcd}(24,n)$, and hence the existence of a theory saturating divisibility is obvious in these cases. 

On the other hand, for $n = 6, 9, 14, 22$ the gcd is larger than $24 / \mathrm{gcd}(24,n)$.  However, notice that an element with $n = 24$ is realized and that at least one of $\mathfrak{I}_n$ and $\mathfrak{I}_{24-n}$ saturates divisibility for all $n$.  Suppose $\mathfrak{I}_n \neq 24/\gcd(24,n)$. Then we can put the theory with $c_\text{L} = 12(24-n)$ and Witten index $\mathfrak{I}_{24-n} = 24/\gcd(24,n)$ on the right (so that it is anti-holomorphic with $c_\text{R} = 12(24-n)$), and take the tensor product with the holomorphic theory realizing $\mathfrak{I}_{24} = 1$.  The result is a (no longer holomorphic) theory with $\nu = -24n$ and Witten index $\mathfrak{I}_{24-n} = 24/\gcd(24,n)$.

\begin{table}[!t]
    \centering
    \begin{tabular}{|c|ccccccccccccc|}
        \hline
        $n$ & 0 & 1 & 2 & 3 & 4 & 5 & 6 & 7 & 8 & 9 & 10 & 11 & 12 \\\hline
        $\mathfrak{I}_n$ & 1 & 24 & 12 & 8 & 6 & 24 & \cellcolor{gray!25}8 & 24 & 3 & \cellcolor{gray!25}24 & 12 & 24 & 2 \\
        $24/\gcd(24,n)$ & 1 & 24 & 12 & 8 & 6 & 24 & \cellcolor{gray!25}4 & 24 & 3 &\cellcolor{gray!25} 8 & 12 & 24 & 2 \\\hline\hline
        $n$ & 24 & 23 & 22 & 21 & 20 & 19 & 18 & 17 & 16 & 15 & 14 & 13 & \\\hline
        $\mathfrak{I}_n$ & 1 & 24 & \cellcolor{gray!25}24 & 8 & 6 & 24 & 4 & 24 & 3 & 8 & \cellcolor{gray!25}24 & 24 & \\
        $24/\gcd(24,n)$ & 1 & 24 & \cellcolor{gray!25}12 & 8 & 6 & 24 & 4 & 24 & 3 & 8 & \cellcolor{gray!25}12 & 24 & \\\hline
    \end{tabular}
    \caption{Values of $\mathfrak{I}_n$ and $24/\gcd(24,n)$ for gravitational anomaly $\nu = -24n$ for $n = 0, 1, \dotsc, 24$.  The cases where $\mathfrak{I}_n$ differ from $24/\gcd(24,n)$ are shaded.}
    \label{Tab:SaturationNaive}
\end{table}

\section*{Acknowledgements}

We are grateful to Kantaro Ohmori, Du Pei, Pavel Putrov, Brandon C.~Rayhaun, Xi Yin, Yunqin Zheng, and especially Theo Johnson-Freyd and Yuji Tachikawa for helpful discussions.
We also thank Theo Johnson-Freyd and an anonymous PTEP referee for comments on the draft.
JA and YL thank Porto University and the Bootstrap 2022 Yearly Collaboration Workshop for hospitality during the progress of this work.
JA is supported in part by NSF grant PHY-2210533.
YL is supported by the Simons Collaboration Grant on the Non-Perturbative Bootstrap.

\appendix

\section{Permutation anomaly}
\label{Sec:Permutation}

In this appendix, we discuss the potential anomalies of the permutation symmetry $\S_n$ of the $n$-th tensor product of a theory $\cT$. Such anomalies can occur when $\T$ has a global gravitational anomaly \cite{Johnson-Freyd:2017ble}.  
If $\T$ has a global symmetry $G$, there can also be mixed anomalies between $\S_n$ and $G^n$, including the diagonal $G$.  
As we will now describe, the pure \emph{cyclic} permutation anomaly can be computed by examining the spins in the $g$-twisted defect Hilbert space, assuming that $\T$ is a holomorphic CFT with $\nu = 2c$ units of the gravitational anomaly.\footnote{The assumption of holomorphy is just so that we can rightfully ignore the energies of states, and only keep track of the Lorentz spins.}
Considering cyclic subgroups of $\S_n$ allows us to derive necessary conditions for the $\S_n$ to be non-anomalous.  
The conditions we derive will also turn out to be sufficient.\footnote{For a general group, it is untrue that the anomalies of cyclic subgroups determine the anomaly of the full group. For $\S_n$ though, we can see that the conditions are sufficient by comparing our results to the rigorous cohomological results. In the bosonic case, this rigorous result can be found in \cite[Theorem 2]{Evans:2018qgz}. For the fermionic case, we may proceed as follows. First note that permutation anomalies can only depend on the gravitational anomaly of $\cT$, or in other words, on the central charge $c\in \half \ZZ$. Moreover, the dependence is linear. Using the fact that $\text{SH}^3(\text{S}^n) = \ZZ_{24}$ when $ n \geq 4$, we then conclude that the orbifold is non-anomalous when $\nu = 2c\in 24\ZZ$. This may be compared to the results around (\ref{eq:fermanom}), and confirms the sufficiency of our conditions. We thank Theo Johnson-Freyd for explaining this to us. 
}

\paragraph{Bosonic case:}  We begin with the bosonic (non-spin) case, where $\nu = 16\mu$ and $\mu \in \bZ$.  
Let $\rho$ be a generator of the $\CC_n$ cyclic permutation symmetry of $\cT^{\otimes n}$.  Then the torus partition function of $\T^{\otimes n}$ twisted by $\rho$ in the temporal direction is
\ie\label{BosonicPermutationTrace}
    Z[\cT^{\otimes n}]^\rho(\tau) &= \Tr_{\cH^{\otimes n}} \rho\, q^{\sum_{i=1}^n L_0^{(i)}-\myfrac{n\mu}{3}} 
    \\
    &= \sum_{\psi_1, \dots, \psi_n \in \cH} \langle \psi_2, \dots, \psi_n, \psi_1 | \psi_1, \dots, \psi_n \rangle\, q^{\sum_{i=1}^n s_i-\myfrac{n\mu}{3}}
    \\
    &= \sum_{\psi_1 \in \cH} q^{n s_1-\myfrac{n\mu}{3}} = Z[\cT](n\tau)~.
\fe
An intuitive way to arrive at the above answer is to regard $\T^{\otimes n}$ as $\T$ living on an $n$-sheeted cover of a flat torus with the same complex modulus $\tau$, with $\rho$ providing branch cuts connecting the $n$ sheets.  A $\rho$-twist in the temporal direction means that the branch cuts are extended in the spatial direction, so the $n$ sheets are woven into a flat torus of complex modulus $n\tau$.  This perspective will be very useful when considering the fermionic case.

By performing a modular S transform and using the modular invariance of $Z[\cT]$, we obtain the torus partition function of $\T^{\otimes n}$ twisted by $\rho$ in the spatial direction,
\ie
    Z[T^{\otimes n}]_\rho(\tau) = Z[T]\left(\myfrac{\tau}{n}\right) = q^{-\myfrac{\mu}{3n}} +\cdots,
\fe
whose Fourier powers are in $\myfrac{\Z}{n} - \myfrac{\mu}{3n}$.
Under the state-operator map, and with the overall conformal factor $q^{\myfrac{n\mu}{3}}$ taken into account, the spins of operators in the $\rho$-twisted defect Hilbert space of $\T^{\otimes n}$ are 
\ie
    s \in \myfrac{\mu(n^2-1)}{3n} + \myfrac{\Z}{n}~.
\fe
This can be compared with the general spin content in the defect Hilbert space of a $\Z_n$ global symmetry with $k \pmod{n}$ units of the anomaly,
\ie
    s \in \myfrac{k}{n^2} + \myfrac{\Z}{n}~,
\fe
and we find that the cyclic permutation $\rho$ has an anomaly
\ie
    k = \myfrac{\mu(n^2-1)n}{3} \mod{n}~,
\fe
which vanishes when $3 \dv \mu(n^2-1)$.  
When $3 \dv \mu$, i.e.\ when the global gravitational anomaly vanishes, $\CC_n$ is non-anomalous for all $n$ (see \cite{Bischoff:2018loc} for a similar result).
It turns out that $\S_n$ for any $n>2$ is non-anomalous if and only if $3 \dv \mu$; see \cite[Theorem 2]{Evans:2018qgz}.

\paragraph{Fermionic case:}  For fermionic (spin) CFTs, we repeat the above exercise while carefully keeping track of the spin structure and statistics.  A nice discussion on fermionic anomalies and their allowed spins in defect Hilbert spaces can be found in \cite{Grigoletto:2021zyv}.  Consider $\T$ on an $n$-sheeted flat torus with complex modulus $\tau$ and Neveu-Schwarz (NS) boundary conditions in both directions.  By inserting $\rho$ branch cuts, we obtain a flat torus with complex modulus $n\tau$ with NS boundary condition in the spatial direction, but the boundary condition in the temporal direction is either Ramond (R) or NS depending on whether $n$ is even or odd,
\ie 
    Z[\cT^{\otimes n}]^{\NS,\rho}_\NS(\tau) = \begin{cases}
        Z[\cT]^\NS_\NS(n\tau) & n \in 2\bN-1~,
        \\
        Z[\cT]^\R_\NS(n\tau) & n \in 2\bN~.
    \end{cases}
\fe 
Under a modular S transformation,
\ie 
    Z[\cT^{\otimes n}]_{\NS,\rho}^\NS(\tau) = \begin{cases}
        \displaystyle
        Z[\cT]_\NS^\NS\left(\myfrac{\tau}{n}\right) = q^{-\myfrac{\nu}{48n}} + \dots & n \in 2\bN-1~,
        \\
        \vspace{-0.15in}
        \\
        \displaystyle
        Z[\cT]_{\R}^\NS\left(\myfrac{\tau}{n}\right) = q^{\myfrac{\ER}{n}} + \dots & n \in 2\bN~,
    \end{cases}
\fe 
where $\ER = \myfrac{\nu}{24} \mod{1}$ is the energy of the R sector ground state of $\T$ on the cylinder.
Under the state-operator map, and with the overall NS conformal factor $q^{\myfrac{n\nu}{48}}$ taken into account, the spins of operators in the $\sigma$-twisted NS defect Hilbert space of $\T^{\otimes n}$ are 
\ie\label{FermionicSpinSelection}
    s \in \begin{cases}
        \displaystyle
        \frac{\nu(n^2-1)}{48n} + \frac{\bZ}{2n} & n \in 2\bN-1~,
        \\
        \vspace{-0.15in}
        \\
        \displaystyle
        \frac{\nu(n^2+2)}{48n} + \frac{\bZ}{n} & n \in 2\bN~.
    \end{cases}
\fe 

Now consider $n=2$.  In this case, the fermionic $\bZ_2$ anomaly has a $\bZ_8$ classification.  We compare \eqref{FermionicSpinSelection} with the general spin content in the NS defect Hilbert space twisted by a $\bZ_2$ global symmetry with $k \mod{8}$ units of the anomaly\footnote{See for instance \cite[(5.8)]{Grigoletto:2021zyv}.}
\ie
    s \in \myfrac{k}{16} + \myfrac{\bZ}{2}~,
\fe
and find that $\sigma$ has
\ie
    k = \nu \mod{8}~,
\fe
which vanishes if and only if $8 \dv \nu$.

Next consider prime $n>2$.  In this case, the fermionic $\bZ_n$ anomaly has a $\bZ_n$ classification.  We compare \eqref{FermionicSpinSelection} with the general spin content with $k \pmod{n}$ units of the anomaly
\ie
    s \in \myfrac{k}{2n^2} + \myfrac{\Z}{2n}~,
\fe
to find that $\sigma$ has
\ie
\label{eq:fermanom}
    k = \myfrac{\nu n(n^2-1)}{24} \mod{n}~,
\fe
which always vanishes for $n>3$, and vanishes for $n=3$ if and only if $3 \dv \nu$.

When $24 \dv \nu$, both $\CC_2$ and $\CC_3$ are  non-anomalous.
It turns out that $\S_n$ for any $n>2$ is non-anomalous if and only if $24 \dv \nu$.

\section{Proofs of alternating orbifold formulae}

\label{app:Proofs}

\subsection{Proof of Theorem \ref{thm:AltDMVV} }
\label{app:altDMVVproof}
In this appendix, we derive the analog of the DMVV formula for alternating orbifolds, given in Theorem \ref{thm:AltDMVV}.  Before giving the proof, we first give a bit of background. Recall that the Hilbert space of an orbifold theory can be written as a sum over conjugacy classes $[h]$ of twisted sectors $\Hilb_h$ wherein one projects onto states invariant under the centralizer subgroup $\cen_h$,
\begin{equation}\label{eq:orbifoldH}
    \mathcal{H}\left(\T^{\otimes n}/\S_n\right)=\bigoplus_{[h]}\Hilb(\T^{\otimes n})_h^{\cen_h}~,
\end{equation}
so that \eqref{eq:SnOrb} can be rewritten as
$Z[\T^{\otimes n}/\S_n] = \sum_{[h]} 
    Z[\T^{\otimes n}]^{\cen_h}_h~.$
The group elements $h\in \S_n$ are permutations of $n$ elements and can be decomposed into disjoint cycles $(k)$ of $k$ elements as
\begin{equation}\label{eq:cycledecomp}
    h=(1)^{n_1}(2)^{n_2}\ldots(n)^{n_n}~,
\end{equation}
where $n_k$ indicates the number of cycles of length $k$.\footnote{In this cycle decomposition, all the permuted elements from 1 to $n$ should appear exactly once, so invariant elements should be counted as 1-cycles, e.g.\ $(345)=(1)(2)(345)$.} Since the cycle type is invariant under permutations of the $n$ elements, conjugacy classes $[h]$ are in one-to-one correspondence with partitions $\{n_k\}$ of $n$, i.e.
\begin{equation}
    \sum_k k\, n_k=n~.
\end{equation}
Meanwhile, the centralizer subgroup of $h$ is given by
\begin{equation}\label{eq:centralizer}
    \cen_h = \S_{n_1}\times \left(\S_{n_2}\rtimes \Z_2^{n_2}\right)\times\ldots \times\left(\S_{n_n}\rtimes \Z_n^{n_n}\right)~,
\end{equation}
where the factors $\S_{n_k}$ permute the different cycles of length $k$ and each $\Z_k$ acts by shifting all the elements inside the corresponding $(k)$-cycle. 

It thus follows that
\begin{equation}\label{eq:Horb}
    \mathcal{H}\left(\T^{\otimes n}/\S_n\right)
    =\bigoplus_{\{n_k\}}\Hilb(\T^{\otimes n})_h^{\cen_h}
    =\bigoplus_{\{n_k\}}\bigotimes_{k>0} \S^{n_k}\Hilb(\T^{\otimes k})_{(k)}^{\Z_k}~,
\end{equation}
where $\Hilb(\T^{\otimes k})_{(k)}^{\Z_k}$ is a smaller Hilbert space twisted by the cycle $(k)$ and projected onto the $\Z_k$-invariant states, and $\S^{n_k}$ denotes a symmetric tensor product of $n_k$ copies of it. We will drop the argument $\T^{\otimes k}$ in what follows to reduce clutter. One can now derive the DMVV formula \eqref{eq:DMVV} by computing the elliptic genus of each of the pieces in \eqref{eq:Horb} and making repeated use of the identities $Z[\Hilb_1\oplus \Hilb_2]=Z[\Hilb_1]+Z[\Hilb_2]$ and $Z[\Hilb_1\otimes \Hilb_2]=Z[\Hilb_1]Z[\Hilb_2]$.

As discussed in Section \ref{sec:AltDMVV}, alternating orbifolds can be obtained from their symmetric counterpart by projecting out the contributions from odd permutations (and multiplying by two). For the spatial twists, this implies that we should only keep even partitions of $n$ in the sum of \eqref{eq:Horb}, where the parity of a partition is given by $|h|=\sum_k (k+1)n_k \mod\,2$. For the temporal twists, we insert the projector $\frac{1}{2}(1+\sgn\,g)$ when projecting onto the $\cen_h$-invariant states. The factor $\sgn\,g$ has two effects, in parallel with the decomposition \eqref{eq:centralizer} of the centralizer. First, the transpositions $x_k\in \S_{n_k}$ of two cycles of length $k$ pick up a factor of $(-1)^k$ and thus the symmetric products $\S^{n_k}$ for odd $k$ become \textit{antisymmetric} products $\Lambda^{n_k}$. Second, the generator $\omega\in \Z_k$---which is the cycle $(k)$ itself---gets multiplied by $(-1)^{k+1}$ and therefore instead of projecting onto $\Z_k$-invariant states, we project onto the sector of ``$\Z_k$-odd states'' $\Hilb^{\Z_k^-}_{(k)}$ when $k$ is even, a concept that we will explain momentarily. Summarizing, the Hilbert space for alternating orbifolds is
\begin{equation}\label{eq:HAnOrb}
    \mathcal H\left(\T^{\otimes n}/\A_n\right)= \bigoplus_{\text{even}\,\{n_k\}}\Bigg( \bigotimes_{k>0} 
    \S^{n_k}\mathcal H_{(k)}^{\mathbb Z_{k}} 
    \:\:\oplus\:\:
    \bigotimes_{k>0}\Lambda^{n_{2k-1}}\mathcal H_{(2k-1)}^{\mathbb Z_{2k-1}} \otimes \S^{n_{2k}}\mathcal H_{(2k)}^{\mathbb Z_{2k}^-}\Bigg)~.
\end{equation}

This is the analog of \eqref{eq:sgn-gauging} at the level of the Hilbert space. Indeed, the first term in \eqref{eq:HAnOrb} corresponds to setting $\beta=0$ in \eqref{eq:Zab} while the second one, where we multiplied by the signature of the temporal twist, is for $\beta=1$. The relation to $\alpha$ comes through the projection onto even partitions of $n$, which is done with $\frac{1}{2}(1+\sgn\,h)$. When multiplying by $\sgn\,h$ (i.e.\ when setting $\alpha=1$) each cycle $(k)$ contributes an overall factor of $(-1)^{k+1}$, adding up to $(-1)^{|h|}$. We can now compute the generating functions $\cZ_{\alpha\beta}$ of \eqref{eq:sgn-gauging} by generalizing the ingredients of the derivation of \eqref{eq:DMVV}.

Following DMVV \cite{DMVV}, we can relate the elliptic genus of $\Hilb_{(k)}$ to $Z[\T]$ with the replacement $\tau\to \tau/k$ since the former can be seen as the Hilbert space of the theory on a torus of period $2\pi n$ in the spatial direction. Then, the projection onto the $\Z_k$-invariant sector is done with the projector $P_k=\frac{1}{k}\sum_b\omega^b$, where $\omega$ implements a $T$ transformation $\tau\to\tau + 1$. In terms of the mode expansion \eqref{eq:Z[T]modes}, we have
\begin{equation}\label{eq:HZn}
    Z[\Hilb_{(k)}^{\Z_k}](\tau,z) =
    \frac{1}{k}\sum_{b=0}^{k-1}\sum_{m,\ell}c(m,\ell)q^{\frac{m}{k}}e^{2\pi i\frac{ b m}{k}}y^\ell=\sum_{m,\ell}c(m k,\ell)q^m y^\ell~,
\end{equation}
where we have used $\sum_b e^{2\pi i\frac{bm}{k}}=k\sum_j\delta_{j,\frac{m}{k}}$. In contrast, when projecting onto the ``$\Z_k$-odd'' sector, we must add a factor $(-1)^{b}$ in the projector $P_k$, so we have $\sum_b (-1)^be^{2\pi i\frac{bm}{k}} = k\sum_j \delta_{j,\frac{m}{k} + \frac{1}{2}}$ and thus only states of half-integer moding contribute,
\begin{equation}\label{eq:HZn-}
    Z[\Hilb_{(k)}^{\Z_k^-}](\tau,z)  =
    \sum_{m,\ell}c((m+\tfrac{1}{2})k,\ell)q^{m+\frac{1}{2}}y^\ell~.
\end{equation}

The next step is to compute the elliptic genus of symmetric and antisymmetric products of $\Hilb_{(k)}^{\Z_k^{\pm}}$. For that, we first consider a generic Hilbert space $\Hilb$ with elliptic genus
\begin{equation}
    Z[\Hilb](\tau,z) = \sum_{m,\ell} d(m,\ell)q^m y^\ell~,
\end{equation}
and we interpret $d(m,\ell)=\text{dim}\,V_{m,\ell}$ as the (super)dimension of a vector space $V_{m,\ell}$. Then the elliptic genera of symmetrized products of $\Hilb$ take a compact form if we consider their generating function \cite{DMVV},
\begin{equation}\label{eq:genfunSn}
    \sum_{n>0}p^n Z[\S^n\Hilb](\tau,z)=\prod_{m,\ell}\sum_{n>0} p^n\left(q^my^\ell\right)^n \text{dim}\left(\S^n V_{m,\ell}\right)~.
\end{equation}
Using that the dimension of the symmetric tensor $\S^n V_{m,\ell}$ is $\text{dim}\left(\S^n V_{m,\ell}\right)=\left(\begin{smallmatrix}d(m,\ell)+n-1\\n\end{smallmatrix}\right)$, we can perform the sum on the right hand side to obtain
\begin{equation}\label{eq:Z[SNH]}
    \sum_{n>0}p^nZ[\S^n\Hilb](\tau,z)=\prod_{m,\ell}\frac{1}{\left(1-pq^my^\ell\right)^{d(m,\ell)}}~.
\end{equation}
For antisymmetric products $\Lambda^n \Hilb$, we reach an expression identical to \eqref{eq:genfunSn} with the replacement $\S^n\to \Lambda^n$. Since the dimension of the antisymmetric tensor $\Lambda^n V_{m,\ell}$ is $\text{dim}\left(\Lambda^n V_{m,\ell}\right)=\left(\begin{smallmatrix}d(m,\ell)\\n\end{smallmatrix}\right)$, we can then also evaluate the sum and get
\begin{equation}\label{eq:Z[LNH]}
    \sum_{n>0}p^nZ[\Lambda^n\Hilb](\tau,z)=\prod_{m,\ell}\left(1+pq^my^\ell\right)^{d(m,\ell)}~.
\end{equation}

We are finally ready to compute the four pieces of \eqref{eq:sgn-gauging}. The form of the symmetric orbifold Hilbert space \eqref{eq:Horb} implies that we can express the generating function of the corresponding elliptic genera as
\begin{align}
    \sum_{n>0}p^n Z[\T^{\otimes n}/\S_n] (\tau,z)=&\,\sum_{n>0}p^n\sum_{\{n_k\}}\prod_{k>0}Z[\S^{n_k} \Hilb_{(k)}^{\mathbb Z_k}](\tau,z)\\
    =&\,\prod_{k>0}\sum_{n>0}p^{kn}Z[\S^n \Hilb_{(k)}^{\mathbb Z_k}](\tau,z)~. \nonumber
\end{align}
When no $\sgn(\cdot)$ line of the quantum symmetry is inserted, plugging \eqref{eq:HZn} and \eqref{eq:Z[SNH]} into the above equation yields the original DMVV formula \eqref{eq:DMVV},
\begin{equation}\label{eq:Z00}
    \mathcal Z_{00}=\prod_{k>0}\sum_{n>0}p^{kn} Z[\S^n \Hilb_{(k)}^{\mathbb Z_k}]
    =\prod_{\begin{smallmatrix} k>0\\m\in\bZ,\ell\end{smallmatrix}} \frac{1}{\left(1-p^k q^m y^\ell\right)^{c(km,\ell)}} ~.
\end{equation}
When inserting a $\sgn(\cdot)$ line along the spatial direction so as to keep track of the signature of the elements $g$, each copy of $\Hilb_{(k)}^{\mathbb Z_k}$ brings in a factor of $(-1)^{k+1}$, so we get instead
\begin{equation}\label{eq:Z10}
    \mathcal Z_{10}=\prod_{k>0}\sum_{n>0}p^{kn}
    (-1)^{n(k+1)}Z[\S^n \Hilb_{(k)}^{\mathbb Z_k}]
    =\prod_{\begin{smallmatrix} k>0\\m\in\bZ,\ell\end{smallmatrix}} \frac{1}{\left(1+(-p)^k q^m y^\ell\right)^{c(km,\ell)}} ~.
\end{equation}
In contrast, when the $\sgn(\cdot)$ line runs along the time direction, we get from (\ref{eq:HZn},\ref{eq:HZn-},\ref{eq:Z[LNH]}),
\begin{align}\label{eq:Z01}
     \mathcal Z_{01}=&\,\prod_{\text{odd}\,k>0}\Big(\sum_{n>0}p^{kn}
     Z[\Lambda^n \Hilb_{(k)}^{\mathbb Z_k}]\Big)
     \prod_{\text{even}\,k>0}\Big(\sum_{n>0}p^{kn}
     Z[\S^n \Hilb_{(k)}^{\mathbb Z_k^-}]\Big)\\
     =&\,\prod_{\begin{smallmatrix}k>0 \\m\in\bZ,\ell\end{smallmatrix}} \frac{\Bigl( 1 + p^{2k-1} q^{m}y^\ell\Bigr)^{c((2k-1)m,\ell)} \hspace{-17mm}}{\Bigl( 1 - p^{2k} q^{m+\frac{1}{2}}y^\ell\Bigr)^{c(k(2m+1),\ell)}\hspace{-17mm}}\hspace{17mm}~.
\end{align}
Adding on top of this a spatial $\sgn(\cdot)$ line replaces again $p^k\to -(-p)^k$, which yields
\begin{align}\label{eq:Z11}
     \mathcal Z_{11}
     =\prod_{\begin{smallmatrix}k>0 \\m\in\bZ,\ell\end{smallmatrix}} \frac{\Bigl( 1 + p^{2k-1} q^{m}y^\ell\Bigr)^{c((2k-1)m,\ell)} \hspace{-17mm}}{\Bigl( 1 + p^{2k} q^{m+\frac{1}{2}}y^\ell\Bigr)^{c(k(2m+1),\ell)}\hspace{-17mm}}\hspace{17mm}~.
\end{align}
By plugging these results into \eqref{eq:sgn-gauging} one arrives at \eqref{eq:AltDMVV}. This concludes the proof of Theorem \ref{thm:AltDMVV}. 
\newline

\subsection{Proof of Theorem \ref{thm:alttor}}
\label{app:alttor}

In this appendix, we prove Theorem \ref{thm:alttor}, namely the formula for alternating orbifolds with discrete torsion. Before doing so, it is useful to first review the calculation for symmetric orbifolds \cite{Dijkgraaf:1999za}. First, we will need an explicit form for the discrete torsion phase $\epsilon(h,g)$, for any element $h\in \S_n$ and $g\in \cen_h$. As discussed in Section \ref{sec:discrete-torsion}, this can be obtained by lifting $g,h$ to the central extension $\widehat \S_n$ and computing the commutator \eqref{eq:commutator}. The group $\widehat \S_n$ is generated by the lift $\hat t_i$ of the transpositions $t_i=(i\;i+1)\in\S_n$ and an element $z$ analogous to $(-1)^F\in \mathrm{Pin}^-(n-1)$, satisfying the relations
\begin{align}
    \hat t_i^2=&\,z\nonumber\\
    \hat t_i\hat t_{i+1}\hat t_i=&\,\hat t_{i+1}\hat t_i\hat t_{i+1}\nonumber\\
    \hat t_i\hat t_j=&\,z\hat t_j\hat t_i~.\qquad (\text{for}\;\; j>i+1)\hspace{-20mm}
\end{align}
It is technically non-trivial but conceptually straightforward to lift any two elements $g,h$ to $\widehat \S_n$ and compute their commutator using these rules \cite{Dijkgraaf:1999za}. Since the centralizer in $\S_n$ factorizes as \eqref{eq:centralizer} and $\epsilon(g,h)$ forms a representation of $\cen_h$, it is enough to compute this phase for two types of generating elements:
\begin{itemize}
	\item[i)] for a generator $\omega$ of $\mathbb Z_k$ (such that $\omega^k=e$),
	\begin{equation}
		\epsilon(\omega,h)=\begin{cases} \;1 & k\;\text{odd}\\
	    \;(-1)^{|h|-1} & k\;\text{even}~,
		\end{cases}
	\end{equation}
	
	\item[ii)] for a transposition $x_k$ that permutes two cycles of length $k$,
	\begin{equation}
		\epsilon(x_k,h)=(-1)^{k-1}~.
	\end{equation}
\end{itemize}
Then to compute the $\S_n$ orbifold with discrete torsion one just has to repeat the derivation of the DMVV formula keeping track of the additional minus signs due to $\epsilon(g,h)$. Originally, the Hilbert space was given by \eqref{eq:Horb}. With the phase $\epsilon(g,h)$ it becomes more involved, but luckily we already computed all the ingredients we will need in Appendix \ref{app:altDMVVproof} . The effect of $\epsilon(g,h)$ shows up in two different steps of the calculation, relating to the two cases above:
\begin{itemize}
	\item[i)] The $\Z_k$-projector in $\Hilb_{(k)}^{\Z_k}$ now becomes $P_k= \frac{1}{k}\sum_b \epsilon(\omega,h)^b\omega^b$, where $h$ is the full element that we twist by. When $h$ is an odd permutation $\epsilon(\omega,h)$ trivializes and we project onto the $\Z_k$-invariant states $\mathcal H_{(k)}^{\mathbb Z_k}$, whereas when $h$ is an even permutation $\epsilon(\omega,h)=(-1)^{k+1}$ and so we keep the $\Z_k$-invariant states for $k=\text{odd}$ but the $\Z_k$-odd states for $k=\text{even}$. The corresponding elliptic genera were computed in \eqref{eq:HZn} and \eqref{eq:HZn-}.
	
	\item[ii)] The second part of the discrete torsion phase is relevant in the calculation of the elliptic genus of $\S^{n_k}\mathcal H$. When $k= \text{odd}$, $\epsilon(x_k,h)$ is trivial and we recover the result \eqref{eq:Z[SNH]}. When $k=\text{even}$, in contrast, we weight the transpositions $x_k$ by a $-1$, defining the antisymmetric product $\Lambda^{n_k}\mathcal H$. The elliptic genera for this case are given by \eqref{eq:Z[LNH]}.
\end{itemize}

\noindent
All in all, the Hilbert space of the $\S_n$ orbifold with discrete torsion is \cite{Dijkgraaf:1999za}
\begin{align}\label{eq:HSNtor}
    \mathcal{H}\left(\T^{\otimes n}/\S_n^{\text{tor}}\right)=&\,\bigoplus_{\text{even}\,\{n_k\}}\bigotimes_{k>0} \S^{n_{2k-1}}
    \mathcal H_{(2k-1)}^{\mathbb Z_{2k-1}} \otimes \Lambda^{n_{2k}}
    \mathcal H_{(2k)}^{\mathbb Z_{2k}^-} \nonumber \\
    &\, \bigoplus_{\text{odd}\,\{n_k\}}\bigotimes_{k>0} \S^{n_{2k-1}}
    \mathcal H_{(2k-1)}^{\mathbb Z_{2k-1}} \otimes \Lambda^{n_{2k}}
    \mathcal H_{(2k)}^{\mathbb Z_{2k}}~.
\end{align}
Now it is straightforward to compute the elliptic genus of each of these terms using the results of Appendix \ref{app:altDMVVproof}. The only subtlety might be how to disentangle the even and odd partitions of $n_k$, but this can be easily done by inserting the projectors $\frac{1}{2}(1\pm\sgn\,h)$ in the sum. Like in the discussion around \eqref{eq:Z10}, the factor $\sgn\,h$ brings in a factor $(-1)^{k+1}$ to the summand whose net effect is to shift $p^k\to -(-p)^k$. This reproduces Dijkgraaf's result \eqref{eq:SNtor} \cite{Dijkgraaf:1999za}.

We can finally come back to alternating orbifolds. As discussed in Section \ref{sec:discrete-torsion}, alternating orbifolds with $\Z_2$ discrete torsion can again be obtained by projecting out the contributions from odd permutations to the calculation above (and multiplying by two). We have to deal both with twists in the temporal and spatial directions. This is easy for the twists in the spatial direction; we just have to throw away the second line in \eqref{eq:HSNtor}. For the temporal twists we insert the projector $\frac{1}{2}(1+\sgn\, g)$ and recall from \eqref{eq:HAnOrb} that the factor $\sgn\, g$ flips the sign of odd permutations in two different places; the projection onto the sector $\Hilb_{(k)}^{\Z_k}\to\Hilb_{(k)}^{\Z_k^-}$ for $k=\text{even}$ and the symmetric products $\S^{n_k}\to\Lambda^{n_k}$ for $k=\text{odd}$. Thus, the Hilbert space for alternating orbifolds with $\Z_2$ discrete torsion is
\begin{equation}\label{eq:HcSN}
    \mathcal H\left(\T^{\otimes n}/\A_n^{\text{tor}}\right)= \bigoplus_{\text{even}\,\{n_k\}}\Bigg( \bigotimes_{k>0}
    \S^{n_{2k-1}}\mathcal H_{(2k-1)}^{\mathbb Z_{2k-1}} \otimes \Lambda^{n_{2k}}\mathcal H_{(2k)}^{\mathbb Z_{2k}^-}
    \:\oplus\:
    \bigotimes_{k>0}\Lambda^{n_{k}}\mathcal H_{(k)}^{\mathbb Z_{k}} \Bigg)~.
\end{equation}
From this point it is completely straightforward to obtain  \eqref{eq:Atorgenfun} with the results of Appendix \ref{app:altDMVVproof}, concluding the proof of Theorem \ref{thm:alttor}.

\subsection{Proof of Theorem \ref{thm:genHeckeA}}
\label{app:Heckeproof}

In this appendix, we prove Theorem \ref{thm:genHeckeA}, namely the formula for the generating function of alternating orbifolds in terms of generalized Hecke operators. The first step is to define operations on the Fourier coefficients of the expressions appearing in (\ref{eq:AltDMVV}). Having done so, we will then repackage the Fourier coefficients into modular orbits for index $m$ congruence subgroups. For simplicity, we will work with the un-twisted/un-twined versions of the formula. Reintroducing the factors of $g,h$ is straightforward. 

We start with the simplest case of $\cZ_{10}$. In this case, we proceed as follows
\bea
\log \left[ \prod_{\substack{n > 0 \\ m \geq 0}} (1+(-p)^n q^m )^{-c_{m n}}\right]  &=& - \sum_{n >0}\sum_{m\geq 0} c_{mn} \log\left(1+ (-p)^n q^m\right)
\no\\
&=&\sum_{n >0}\sum_{m\geq 0} \sum_{t > 0} (-1)^t c_{mn} {(-p)^{nt} q^{mt} \over t}
\no\\
&=&\sum_{n >0}\sum_{m\geq 0} \sum_{\substack{t | (m,n) \\ t>0}} (-1)^{n+t} t^{-1} c_{m n \over t^2} p^n q^m~. \no
\eea
In the final step we redefined $m \rightarrow m/t$ and $n \rightarrow n/t$. 
We then define the following operation 
\bea
\mathsf{T}_n^{(\alpha, 0)}(Z) = \sum_{m\geq 0} \sum_{\substack{t  | (m,n) \\ t>0}} (-1)^{\alpha(n+t)} t^{-1} c_{m n \over t^2}q^m~.
\eea
For $\alpha=0$ we reproduce the usual Hecke operators, whereas for $\alpha=1$ we reproduce the generalized Hecke operation relevant for $\cZ_{10}$. 

To put this in the form of a modular orbit, we now make the following change of summation variables. We first redefine $m= \ell t$, which gives 
 \bea
 \mathsf{T}_n^{(\alpha, 0)}(Z) = \sum_{\ell\geq 0} \sum_{\substack{t  | n \\ t>0}} (-1)^{\alpha(n+t)} t^{-1} c_{n \ell \over t}q^{\ell t}~.
 \eea
 We then define $t=n/d$ to obtain 
 \bea
 \mathsf{T}_n^{(\alpha, 0)}(Z) =n^{-1} \sum_{\ell\geq 0} \sum_{\substack{d  | n \\ d>0}} (-1)^{\alpha(n+{n\over d})} d\,c_{\ell d}\,q^{\ell n \over d}~.
 \eea
Next we replace $\ell$ with a new $m$, now defined as $m = \ell d$,
 \bea
 \mathsf{T}_n^{(\alpha, 0)}(Z) &=&n^{-1} \sum_{m\geq 0} \sum_{\substack{d  | (m,n) \\ d>0}} (-1)^{\alpha(n+{n\over d})} d\,c_m\,q^{n m \over d^2}
 \no\\
 &=& n^{-1} \sum_{\substack{d  | n \\ d>0}}\sum_{b=0}^{d-1} \sum_{m \geq 0} c_m q^{n m \over d^2} e^{2 \pi i {b m \over d}} (-1)^{\alpha(n+{n\over d})}~,
 \eea
where in the second line we have inserted $\sum_{b = 0}^{d-1} e^{2 \pi i b m /d}$, which is equal to $d$ if $d | m$ and is zero otherwise. This may finally be reassembled into a sum of index $n$ subgroups of $SL(2, \ZZ)$, 
\bea
 \mathsf{T}_n^{(\alpha, 0)}(Z) &=& n^{-1}  \sum_{\substack{d  | n \\ d>0}}\sum_{b=0}^{d-1} (-1)^{\alpha(n+{n\over d})}Z\left({n \tau + b d \over d^2} \right)  \no\\
 &=& n^{-1}  \sum_{\substack{ad=n \\ 0 \leq b < d}}(-1)^{\alpha a (d+1)} Z\left({a \tau + b \over d} \right) 
\eea
where in the last step we defined $a = n/d$. This completes the proof for $\beta = 0$. 

We now proceed to the conceptually straightforward but technically more challenging case of $\beta = 1$. For simplicity, we also take $\alpha = 1$. The starting point is 
 \bea
 &\vphantom{,}&\log \left[ \prod_{\substack{n > 0 \\ m \geq 0}} (1+p^{2n-1} q^m )^{c_{(2n-1)m}} (1+p^{2n} q^{m+{1\over 2}} )^{-c_{n(2m+1)}}\right]
 \\ 
 &=&\sum_{n>0} \sum_{m \geq 0 } c_{(2n-1)m} \log(1+p^{2n-1} q^m) - \sum_{n>0} \sum_{m \geq 0 } c_{n(2m+1)} \log(1+p^{2n} q^{m+{1\over 2}}) 
 \no\\
  &=& \sum_{\substack{n>0\\ m \geq 0}} \sum_{t>0} (-1)^{t+1} c_{(2n-1)m} {p^{(2n-1)t} q^{mt} \over t} -\sum_{\substack{n>0\\ m \geq 0}} \sum_{t>0} (-1)^{t+1} c_{n(2m+1)} {p^{2nt} q^{(m+{1\over 2})t} \over t}
  \no\\
    &=& \sum_{\substack{n>0\\ m \geq 0}} \sum_{t|(m,n)} (-1)^{t+1} c_{({2n\over t}-1){m\over t}} {p^{2n-t} q^{m} \over t} -\sum_{\substack{n>0\\ m \geq 0}} \sum_{t|(m,n)} (-1)^{t+1} c_{{n \over t}({2m\over t}+1)} {p^{2n} q^{m+{t\over 2}} \over t}
  \no\\
  &=&  \sum_{\substack{n>0\\ m \geq 0}} \sum_{\substack{t|m  \\ n/t \in 2 \ZZ+1}} (-1)^{t+1} c_{n m \over t^2} {p^{n} q^{m} \over t} -\sum_{\substack{n\in 2 \NN_{>0}\\ m \in \ZZ/2}} \sum_{\substack{t|(2m,n/2)\\ 2m/t \in 2\ZZ+1}} (-1)^{t+1} c_{n m \over t^2} {p^{n} q^{m} \over t} 
  \eea
In the second equality we used the Taylor expansion of the logarithm. In the third equality we switched summation variables from $m \rightarrow m/t$, $n \rightarrow n/t$. Finally, in the last equation we redefined $2n - t \rightarrow n$ in the first sum and $m+{t\over 2} \rightarrow m $, $2n \rightarrow n$ in the second sum, being careful to introduce the correct constraints on all of the sums. We may write the final result in a marginally more streamlined form as 
\bea
 \sum_{\substack{n>0\\ m \in \ZZ/2}} \sum_{\substack{t\in \NN_{>0}}} (-1)^{t+1} c_{n m \over t^2} {p^{n} q^{m} \over t} \left[\delta_{t|m, {n \over t} \in 2 \ZZ+1, m \in \ZZ} - \delta_{t | (2m,{n\over2}) , n \in 2 \ZZ, {2m \over t} \in 2 \ZZ+1}  \right]
 \eea
where the delta functions impose the relevant constraints on summation variables. We then define the generalized Hecke operator via its action on the Fourier coefficients, 
\bea
\mathsf{T}^{(1,1)}_n(Z) =  \sum_{\substack{m \in \ZZ/2}} \sum_{\substack{t\in \NN_{>0}}} (-1)^{t+1} c_{n m \over t^2} {q^{m} \over t} \left[\delta_{t|m, {n \over t} \in 2 \ZZ+1, m \in \ZZ} - \delta_{t | (2m,{n\over2}) , n \in 2 \ZZ, {2m \over t} \in 2 \ZZ+1}  \right]~.\no
\eea
We now want to reexpress this as a sum over modular orbits. To do so, we follow the same steps as for $\beta = 0$. We begin by swapping the summation variable $m$ for a summation variable $\ell$ defined by $m = \ell t$. We then replace $t$ with $d$ defined by $t=n/d$. This gives 
\bea
\mathsf{T}^{(1,1)}_n(Z) = n^{-1}\sum_{d | n} \sum_{\ell \in \ZZ/2} c_{\ell d} q^{\ell n \over d} \,d\,(-1)^{{n \over d}+1}\left[\delta_{d \in 2 \ZZ+1, \ell \in \ZZ} - \delta_{d \in 2 \ZZ, n \in 2 \ZZ, 2 \ell \in 2 \ZZ+1} \right]~. 
\no
\eea
Next we introduce a new $m$ defined as $m = \ell d$, giving 
\bea
\mathsf{T}^{(1,1)}_n(Z) =n^{-1}\sum_{d | n} \sum_{m \in {d\over 2}\ZZ} c_{m} q^{n m \over d^2} \,d\,(-1)^{{n \over d}+1}\left[\delta_{d \in 2 \ZZ+1, d|m} - \delta_{d \in 2 \ZZ, n \in 2 \ZZ,  {2m \over d} \in 2 \ZZ+1} \right]~. 
\no
\eea
By inserting a factor of ${1\over d}\sum_{b = 0}^{d-1} e^{2 \pi i b m /d}$, the first term on the right-hand side becomes 
\bea
n^{-1} \sum_{\substack{d|n\\ d \in 2 \ZZ+1} } \sum_{b=0}^{d-1} (-1)^{{n \over d}+1} Z \left( {n \tau + b d \over d^2}\right) = n^{-1} \sum_{\substack{ad=n \\ d\in 2 \ZZ+1}}\sum_{b=0}^{d-1} (-1)^{a+1}Z \left( {a \tau + b \over d}\right) ~.
\no
\eea
On the other hand, for the second term on the right-hand side we want to impose ${2m \over d} \in 2 \ZZ+1$ instead of $d|m$, and hence we should insert a factor of ${1\over d}\sum_{b = 0}^{d-1} e^{2 \pi i b (m-{d\over 2}) /d}$, from which we obtain 
\bea
n^{-1} \sum_{\substack{d|n\\ d \in 2 \ZZ} } \sum_{b=0}^{d-1} (-1)^{b+{n \over d}+1} Z \left( {n \tau + b d \over d^2}\right) \delta_{n \in 2 \ZZ} = n^{-1} \sum_{\substack{ad=n \\ d\in 2 \ZZ}}\sum_{b=0}^{d-1} (-1)^{a+b+1}Z\left( {a \tau + b \over d}\right) 
\no
\eea
Adding the two pieces, we obtain the remarkably simple formula 
\bea
\mathsf{T}^{(1,1)}_n(Z) = n^{-1} \sum_{\substack{ad = n \\ 0 \leq b < d} } (-1)^{a+1+(b+1)(d+1)} Z\left( {a \tau + b \over d}\right) ~. 
\eea
Combining this with the result for $\mathsf{T}^{(\alpha,0)}_m(Z)$ given above proves the formula (\ref{eq:genHeckeA}).

\bibliography{refs}
\bibliographystyle{JHEP}

\end{document}